\documentclass[a4paper,11pt]{article}
\pdfoutput=1
\usepackage{jcappub}

\usepackage{mathtools}
\usepackage{fontawesome} 

\newcommand{\rR}{\rho_R}
\newcommand{\gs}{g_\star}
\newcommand{\gss}{g_{\star s}}
\newcommand{\Tev}{T_\text{ev}}
\newcommand{\Mbh}{M_\text{BH}}
\newcommand{\equaref}[1]{Eq.~(\ref{#1})}

\newcommand{\equassref}[3]{Eqs.~(\ref{#1}), (\ref{#2})~and~(\ref{#3})}

\newcommand{\figref}[1]{Fig.~\ref{#1}}

\newcommand{\secref}[1]{Section~\ref{#1}}

\title{Rescuing High-Scale Leptogenesis using Primordial Black Holes}

\author[a]{Nicolás Bernal,}
\author[b]{Chee Sheng Fong,}
\author[c]{\\Yuber F. Perez-Gonzalez}
\author[c]{and Jessica Turner}
\affiliation[a]{Centro de Investigaciones, Universidad Antonio Nariño\\Carrera 3 este \# 47A-15, Bogotá, Colombia}
\affiliation[b]{Centro de Ciências Naturais e Humanas
Universidade Federal do ABC\\09.210-170, Santo André, SP, Brazil}
\affiliation[c]{Institute for Particle Physics Phenomenology, Durham University\\South Road, Durham, U.K.}

\emailAdd{nicolas.bernal@uan.edu.co}
\emailAdd{sheng.fong@ufabc.edu.br}
\emailAdd{yuber.f.perez-gonzalez@durham.ac.uk}
\emailAdd{jessica.turner@durham.ac.uk}

\abstract{We explore the interplay between 
light primordial black holes (PBH) and high-scale baryogenesis, with a particular emphasis on leptogenesis. We first review a generic baryogenesis scenario where a heavy particle, $X$, with mass, $M_X$, produced solely from PBH evaporation  decays to generate a baryon asymmetry. We show that the viable parameter space is bounded from above by $M_X \lesssim 10^{17}$~GeV and increases with decreasing $M_X$. We demonstrate that regions of the leptogenesis parameter space, where the lightest right-handed neutrino (RHN) mass $M_{N_{1}}\gtrsim 10^{15}\,{\rm GeV}$ and neutrino mass scale $m_\nu\gtrsim 0.1$~eV, excluded in standard cosmology due to  $\Delta L=2$ washout processes, becomes viable with the assistance of light PBHs.
This scenario of PBH-assisted leptogenesis occurs because the PBHs radiate RHNs via Hawking evaporation late in the Universe's evolution when the temperature of the thermal plasma is low relative to the RHN mass. Subsequently, these RHNs can decay and produce a lepton asymmetry while the washout processes are suppressed.}

\begin{document}
\begin{flushright}
  PI/UAN-2022-711FT, IPPP/22/12
\end{flushright}

\maketitle

\section{Introduction}\label{sec:intro}

Baryogenesis via leptogenesis is an unavoidable consequence of $SO(10)$ Grand Unified Theories (GUTs), which predict the existence of very heavy right-handed neutrinos (RHN), $N$, with mass-scale $M_N\gtrsim 10^{10}$~GeV~\cite{DiBari:2008mp, Fong:2014gea}. While direct production of such heavy RHNs is not feasible in current or foreseeable future experiments, one can have indirect hints that high-scale leptogenesis occurred in nature. For instance, a particular $SO(10)$ symmetry breaking chain can be correlated with the proton lifetime and its associated gravitational wave (GW) signatures~\cite{King:2021gmj}. In principle, the observation/non-observation of such as terrestrial and cosmological observables could constrain the RHN mass scale. 
Moreover, another interesting feature of $SO(10)$ GUTs is that leptogenesis is intimately connected with the Standard Model (SM) fermion masses~\cite{Fong:2014gea}, since each generation of fermions is unified within the same multiplet at the high scale. Nonetheless, in standard cosmology, thermal leptogenesis is constrained from above by the perturbativity of the Yukawa couplings and by $\Delta L=2$ scatterings, which can erase the lepton asymmetry. The strength of such washout processes increases with the temperature and the absolute light neutrino mass scale, $m_\nu$. As the absolute mass scale may be measured in the near future, the focus of our paper is the
correlation of $m_\nu$ with high-scale leptogenesis and how this parameter space may be enlarged due to the presence of primordial black holes (PBH). 

\begin{figure}
	\centering
 \includegraphics[scale=0.8]{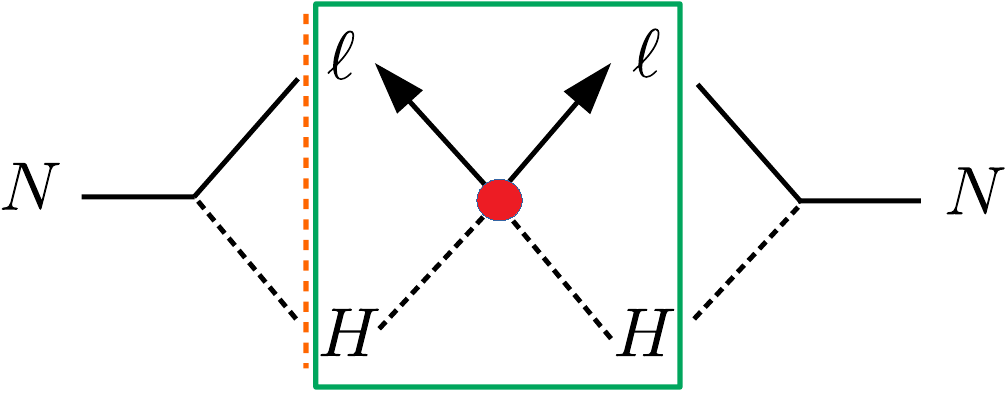}
 \caption{The leading CP violation comes from the interference between the tree-level and one-loop level decay diagrams of $N \to \ell\, H$. The orange dashed line denotes the contribution when the loop particles go on their mass shell. The green square highlights the contribution to light neutrino mass as well as $\Delta L =2$ scattering process for high-scale thermal leptogenesis.}
	\label{fig:CP_mnu_scatt}
\end{figure}

The connection between $m_\nu$ and type-I thermal leptogenesis can be understood as follows. The leading CP-violating interactions of leptogenesis arise from the interference between tree-level and one-loop decay diagrams of $N \to \ell\, H$. This interference is schematically shown in \figref{fig:CP_mnu_scatt} where the dashed orange line denotes the contribution when the leptonic and Higgs doublets ($\ell$ and $H$, respectively) go on their mass shells. The green square highlights the operator that violates lepton number by two units and contributes to the neutrino mass, $m_\nu$, and the lepton-number-violating scattering amplitude e.g. 
$\bar \ell H^\dagger \leftrightarrow \ell H$ and $\ell \ell \leftrightarrow H^\dagger H^\dagger$. The red dot represents heavier degrees of freedom at a scale $\Lambda > M_N$ such as heavier generations of $N$ and/or heavy Higgs $SU(2)_L$ triplet. Schematically, we can write the dimensionless CP violation parameter, $\epsilon$, as 
\begin{equation}
 \frac{\epsilon}{M_N} \propto \frac{m_\nu}{v^2} \propto \sqrt{\sigma_{\Delta L = 2}}\,,\label{eq:general_connection}
\end{equation}
where $v \simeq 174$~GeV is the SM Higgs vacuum expectation value and $\sigma_{\Delta L = 2}$ is the cross-section for scattering processes which violate lepton number by two units. In the above relation, we have assumed that the main contribution to the neutrino mass comes from the operator in the green square of \figref{fig:CP_mnu_scatt}, and we have estimated the cross-section in the regime $T \ll \Lambda$, with $T$ the temperature of the SM thermal bath.
The first relation in \equaref{eq:general_connection} leads to an upper limit on $\epsilon$ which depends on $m_\nu$ and $M_N$ and is known as the Davidson-Ibarra bound~\cite{Davidson:2002qv}.\footnote{A careful derivation in Ref.~\cite{Davidson:2002qv} for type-I seesaw with hierarchical spectrum of $N$ obtained $\epsilon$ which is bounded from above by $m_h - m_l$ with $m_h (m_l)$ is the heaviest (lightest) light neutrino mass.} Imposing a sufficiently large CP-violating parameter, $|\epsilon| \gtrsim 10^{-6}$, implies a lower bound on $M_N \gtrsim (0.1\,{\rm eV}/m_\nu) \times 10^{10}$~GeV where we have included a one loop factor of $1/(16\pi)$ when relating $\epsilon$ to $m_\nu$.

The $\Delta L = 2$ scattering processes in \equaref{eq:general_connection} become important when their rate exceeds the expansion rate of the Universe
\begin{equation}
 \sigma_{\Delta L =2}\, \frac{T^3}{\pi^2} \sim \frac{m_\nu^2}{\pi^3\, v^4}\, T^3 > H(T) \simeq 1.66\sqrt{\gs}\, \frac{T^2}{M_P}\,,
\end{equation}
where we have included a phase space factor of $1/\pi$ for $\sigma_{\Delta L =2}$, and $H(T)$ corresponds to the Hubble expansion rate in a radiation dominated Universe with effective relativistic degrees of freedom $\gs$, and the Planck mass is $M_P \simeq 1.22 \times 10^{19}$~GeV. This condition translates to $T \gtrsim 4\, (0.1\,{\rm eV}/m_\nu)^2 \times 10^{12}$~GeV for $\gs = 106.75$. While this relation is valid only up to $T \sim \Lambda$, it allows us to estimate when $\Delta L = 2$ scattering processes become relevant. 
Thermal leptogenesis suffers from strong washout from $\Delta L = 2$ scattering processes for $M_N \sim T \gtrsim 4\, (0.1\,{\rm eV}/m_\nu)^2 \times 10^{12}$~GeV which will lead to upper bounds on both $M_N$ and $m_\nu$.
The upper bound $m_\nu \lesssim 0.1$~eV has been derived previously such that $\Delta L = 2$ processes do not suppress the generated asymmetry~\cite{Buchmuller:2004nz,Giudice:2003jh}. While cosmological bounds give approximately the same bound~\cite{Ivanov:2019hqk}, nonstandard cosmology \cite{Allahverdi:2020bys} can still allow for large $m_\nu \sim \mathcal{O}(1)$~eV~\cite{Alvey:2021xmq}. 

A possible way to evade the strong washout from $\Delta L = 2$ scatterings consists of generating a population of RHNs after those processes have frozen out. 
Moreover, some additional, non-thermal physical processes should intervene in the RHN production since the Yukawa interactions control both the RHN generation and washout in the standard scenario.
A minimal framework that bypasses the requirement of additional degrees of freedom and/or interactions is particle production via Hawking radiation of PBHs~\cite{Hawking:1975vcx}. 

Baryogenesis via PBH evaporation has attracted substantial attention since the discovery that BHs produce particles with a thermal spectrum~\cite{Toussaint:1978br, Hawking:1980ng, Barrow:1990he, Majumdar:1995yr, Bugaev:2001xr, Baumann:2007yr, Fujita:2014hha, Morrison:2018xla, Ambrosone:2021lsx, Hooper:2020otu}.
Moreover, recent works have also considered the effects of evaporation in leptogenesis~\cite{Perez-Gonzalez:2020vnz, Datta:2020bht, JyotiDas:2021shi, Barman:2021ost}.
The interplay between thermal and PBH-assisted leptogenesis was discussed in detail in Ref.~\cite{Perez-Gonzalez:2020vnz}. This work demonstrated that the PBH-produced RHNs can decay and produce a lepton asymmetry but that the entropy injection from the PBHs can dilute the lepton asymmetry, generated both non-thermally from the PBHs and thermally from the plasma, as PBHs are much more efficient at producing photons rather than RHNs. This dilutionary effect is most severe if the PBHs' mass exceeds $\sim \mathcal{O}(10^3)$~g. In this scenario, the baryon asymmetry obtained in the intermediate-scale thermal leptogenesis scenario (with the lightest RHN masses $10^6 \lesssim M_{N_1}\,({\rm GeV}) \lesssim 10^{8}$) would be diluted even in the most finely tuned regions of the parameter space. 

In this work, we first explore generic baryogenesis where heavy particles $X$ produced purely from PBH evaporation with their subsequent $B-L$-violating and $CP$-violating decays, which generate a cosmic baryon asymmetry. Then, we study how light PBHs with masses smaller than $\mathcal{O}(1)$~g allow the evasion of the bound of $m_\nu \lesssim 0.1$~eV and open up the parameter space of very high-scale leptogenesis up to the GUT scale. The basic idea is that this population of light PBHs can evaporate at sufficiently high temperatures to produce a non-negligible abundance of RHNs when the ambient temperature of the SM plasma is sufficiently low such that the $\Delta L = 2$ scatterings are out of thermal equilibrium. Hence, the rapid decays of these PBH-produced RHNs generate a $B-L$ asymmetry when the washout processes are suppressed.%
\footnote{It is interesting to note that particles (and in particular RHNs) heavier than the reheating temperature could still be produced in the early Universe during the reheating era if their mass is smaller than the highest temperature $T_\text{max}$ reached by the thermal bath~\cite{Giudice:2000ex}. This corresponds to another viable scenario for leptogenesis~\cite{Bernal:2021kaj}.}
As a result, the upper bound on $m_\nu$, which arises due to these scatterings, can be evaded. The direct/terrestrial measurement of the upper bound on the heaviest active neutrino mass comes from the endpoint measurement in the beta-decay spectrum and is around 0.8 eV~\cite{KATRIN:2021uub}. If subsequent measurements determine neutrino masses to be large ($\gtrsim 0.1$ eV)  then much of the parameter space of standard high-scale thermal leptogenesis  would be excluded. However, if light PBHs once existed as a sizable fraction of the early Universe's energy budget, high-scale leptogenesis can still be a viable explanation for the observed matter-antimatter asymmetry.  

The paper is organized as follows: in \secref{sec:leptoPBH} we begin with a general discussion of PBH formation and evaporation as well as considering a generic PBH-assisted baryogenesis scenario. In \secref{sec:lept} we summarize the relevant features of high-scale thermal leptogenesis and $\Delta L =2$ washout processes and present the Boltzmann equations we numerically solve to study the interplay between very high-scale thermal leptogenesis and light PBHs. The main results of the paper are provided in \secref{sec:results}, and we summarize and conclude in \secref{sec:conclusions}. We use natural units in which $\hbar = c = k_{\rm B} = 1$ throughout this manuscript.

\section{Baryogenesis and Primordial Black Holes} \label{sec:leptoPBH}

\subsection{PBH Formation and Evaporation in a Nutshell} \label{sec:PBHnut}
PBHs could have been generated due to large density perturbations in the early Universe~\cite{Hawking:1975vcx, Carr:1974nx} and when these density fluctuations reenter the horizon, they can collapse to form a BH if they exceed a threshold~\cite{Press:1973iz}. This paper focuses on the case where PBHs form in a radiation-dominated epoch at an initial SM radiation temperature $T = T_0$. The initial PBH mass is related to the mass within the particle horizon as~\cite{Carr:2009jm, Carr:2020gox}
\begin{equation}
  M_{\rm BH0} = \frac{4\pi}{3}\, \alpha\, \frac{\rR(T_0)}{H^3(T_0)}\,,
\label{eq:PBH_formation}
\end{equation}
where the gravitational collapse factor is $\alpha\sim 0.2$ in a radiation-dominated era, $\rR$ corresponds to the SM radiation energy density, while the Hubble rate is given by
\begin{equation}
 H = \sqrt{\frac{8\pi}{3} \frac{\rR}{M_P^2}} = \sqrt{\frac{4\pi^{3}\, \gs}{45}}\, \frac{T^{2}}{M_P}\,. \label{eq:H_radiation}
\end{equation}
Here and in the following, we assume that the effective relativistic degrees of freedom, $\gs$, is that of the SM $\gs = 106.75$ for a bath temperature much above the weak scale.
From \equaref{eq:PBH_formation}, we can solve for the cosmic temperature, $T_0$, when a PBH of mass, $M_{\rm BH0}$, is formed
\begin{equation}
 T_0 = \frac12 \left(\frac{5}{\gs \pi^3}\right)^\frac14 \sqrt{\frac{3\, \alpha\, M_P^3}{M_{\rm BH0}}} \simeq 4.27 \times 10^{15}~{\rm GeV} \left(\frac{106.75}{\gs}\right)^\frac14 \left(\frac{1~{\rm g}}{M_{\rm BH0}}\right)^\frac12 \left(\frac{\alpha}{0.2}\right)^\frac12 .
\label{eq:T0}
\end{equation}
This implies that after inflation, the thermal bath temperature should reheat at least to $T_0$ to form such a PBH.
A lower bound on the initial PBH mass can be set once the upper bound on the inflationary scale is considered. The limit reported by the Planck collaboration $H_I \leq 2.5 \times 10^{-5}~M_P$~\cite{Planck:2018jri} implies that $M_\text{BH0} \gtrsim 10^{-1}$~g.
Since for gram-scale PBHs, the inflationary scale must be rather large, one may wonder if there are known mechanisms for formation of such light BHs. One such example can be found in Ref.~\cite{Rasanen:2018fom} where they  considered a scenario that leads to PBH masses below $10^6$ g for some specific features of the Higgs potential. Thus, there could exists models that could predict the formation of very light PBHs. However, we remain agnostic regarding the formation mechanism in what follows.
We define the initial PBH energy density over the
radiation energy density as
\begin{equation}
 \beta \equiv \frac{\rho_{{\rm BH}} \left(T_{0}\right)}{\rR\left(T_{0}\right)} = \frac{n_{0}\, M_{\rm BH0}}{\rR\left(T_{0}\right)}\,,
\label{eq:beta}
\end{equation}
where $n_{0}$ is the initial number density of PBHs, which we assume all have the same mass, $M_{\rm BH0}$.\footnote{In this work, we will not consider extended PBH mass functions that could arise if the PBHs are generated from inflationary density fluctuations or cosmological phase transitions, see e.g. Refs.~\cite{Sasaki:2018dmp, Carr:2020xqk} for reviews.}
Depending on the number density of PBHs generated in the early Universe, they could eventually dominate the energy density of the Universe, leading to a non-standard cosmology~\cite{Allahverdi:2020bys}. 

After formation, the PBHs evaporate via the emission of all possible degrees of freedom in nature~\cite{Hawking:1975vcx}.
Throughout this work, we assume that PBHs are in the Schwarzchild phase and are therefore chargeless and spinless.
The horizon radius, $r_{{\rm S}}$, and the temperature
for such a BH, $T_{{\rm BH}}$, are given by
\begin{equation}
r_{{\rm S}} =\frac{2\Mbh}{M_P^{2}} \qquad \text{and} \qquad T_{{\rm BH}} = \frac{M_P^{2}}{8\pi\, \Mbh}\,,
\end{equation}
respectively, with $\Mbh$ denoting the instantaneous BH mass.
The rate of emission per energy interval of a particle of type $i$ with mass $m_i$ (in the limit $m_i \ll T_\text{BH}$) is given by
\begin{equation}
 \frac{d^{2}N_{i}}{dt\, dE} = \frac{g_{i}}{2\pi}\frac{\vartheta_i(\Mbh,E)}{e^{E/T_{{\rm BH}}}-\left(-1\right)^{2s_{i}}}\,,\label{eq:emission_rate}
\end{equation}
where $g_{i}$ and $s_{i}$ denote the number of degrees
of freedom and spin of particle $i$, respectively. 
The spin-dependent factor, $\vartheta_i(\Mbh,E)$, in the Hawking emission rate, known as a graybody factor, describes the probability of a particle reaching spatial infinity.
This graybody factor has an oscillatory behavior and tends towards the geometric optics limit $\vartheta_i(\Mbh,\, E) \longrightarrow 27 r_{\rm S}^2\, E^2/4$ for large values of the energy $E$.
From \equaref{eq:emission_rate}, the mass loss rate of a PBH can be determined as follows~\cite{MacGibbon:1990zk, MacGibbon:1991tj}
\begin{equation} \label{eq:MEq}
 \frac{d\Mbh}{dt} = -\sum_{i} \int_{m_{i}}^{\infty} \frac{d^{2}N_{i}}{dt\, dE}\, E\, dE = - \varepsilon(\Mbh)\, \frac{M_P^4}{\Mbh^2}\,,
\end{equation}
where $\varepsilon(\Mbh)$ contains the information of the degrees of freedom that can be emitted during the evaporation process as a function of the instantaneous BH mass, see, e.g. Ref.~\cite{MacGibbon:1990zk,Cheek:2021odj}.

Several bounds exist in the PBH parameter space spanned by the initial fraction $\beta$ and mass $M_\text{BH0}$~\cite{Carr:2020gox, Carr:2020xqk}. 
We will focus on PBHs that evaporated before Big-Bang Nucleosynthesis (BBN), which have masses $\lesssim 10^9$~g. 
Although there exist constraints on such light PBHs, they are typically model dependent~\cite{Carr:2020gox, Carr:2020xqk, Lunardini:2019zob}. 
Nevertheless, recent constraints have been derived after considering the GWs emitted from the Hawking evaporation.
In particular, a backreaction problem can be avoided if the energy contained in GWs never exceeds that of the background Universe~\cite{Papanikolaou:2020qtd}.
More importantly, a modification of BBN predictions due to the energy density stored in GWs can be avoided if~\cite{Domenech:2020ssp}
\begin{equation} \label{eq:GW}
  \beta \lesssim 1.1 \times 10^{-6} \left(\frac{\alpha}{0.2}\right)^{-\frac12} \left(\frac{M_\text{BH0}}{10^4~\text{g}}\right)^{-\frac{17}{24}}.
\end{equation}

\subsection{PBH-assisted Baryogenesis}\label{sec:lepasst}
In this section, we consider a generic SM singlet particle $X$ with mass, $M_X$, and internal degrees of freedom, $g_X$, that can decay and produce a $B-L$ asymmetry. 
 We analytically estimate the number of $X$ particles emitted by a PBH to determine the maximum baryon asymmetry that could be obtained from PBH evaporation only.
We will omit the graybody factors and assume that the BH is a perfect black body, i.e. $\vartheta_i(\Mbh,E)\to 1$.
However, let us stress that we will include the graybody factors through our numerical treatment as detailed in \secref{sec:lept}.
As the emission of light or massless particles dominates the mass loss rate when $T_{{\rm BH}}\gg m_{i}$, it is an excellent approximation to set $m_{i}=0$, and hence we can obtain
\begin{equation}
 \frac{d\Mbh}{dt} \approx -\frac{\gs\, M_P^{4}}{30720\pi\, \Mbh^{2}}\,.
\end{equation}
For a PBH with initial mass $M_{\rm BH0}$, its lifetime can be estimated as
\begin{equation}
 \tau \approx -\int_{M_{\rm BH0}}^{0} \frac{30720\pi\, \Mbh^{2}}{\gs\, M_P^{4}}\, d\Mbh = \frac{10240\pi\, M_{{\rm BH0}}^{3}}{\gs\, M_P^{4}}\,.
\end{equation}
Assuming the Universe continues to be radiation dominated, we can compute the radiation temperature, $\Tev$, when the PBH population completely evaporates by setting the Hubble rate from \equaref{eq:H_radiation} to be $H=1/(2\tau)$, obtaining
\begin{equation}
 \left.\Tev\right|_{\rm Rad~dom} \simeq \frac{1}{64}\left(\frac{\gs }{5\pi^{5}}\right)^\frac14 \sqrt{\frac{3\, M_P^{5}}{2\, M_{\rm BH0}^{3}}} \simeq 1.22\times10^{10}~{\rm GeV} \left(\frac{\gs}{106.75}\right)^\frac14 \left(\frac{1~{\rm g}}{M_{\rm BH0}}\right)^\frac32.
\label{eq:Tev}
\end{equation}
On the other hand, if PBHs dominate the cosmic energy density before they completely evaporate, one should reevaluate the above estimation of $\Tev$ as it will now depend on the initial abundance of the PBHs.  
Therefore, to determine $\Tev$, we have to solve for $\Tev$ by setting $H=2/(3\tau)$ and replacing the radiation energy density by the PBH energy density given in \equaref{eq:H_radiation} and make the replacement $\rR \to \rho_M = M_{\rm BH0}\, n(\Tev)$ obtaining
\begin{align}
 \left.\Tev\right|_{\rm PBH~dom}&\simeq \frac{1}{128}\left(\frac{\gs\, M_P^{10} }{10\pi^{5}\, M_{\rm BH0}^6\, \beta\, T_0 }\right)^{1/3}\notag\\
 &\simeq 4.11\times 10^{10}~{\rm GeV}
 \left(\frac{\gs}{106.75}\right)^{5/12}
 \left(\frac{10^{-3}}{\beta}\right)^{1/3}
 \left(\frac{0.2}{\alpha}\right)^{1/6}
 \left(\frac{1~{\rm g}}{M_{\rm BH0}}\right)^{11/6}.
\label{eq:TevM}
\end{align}
We emphasize that in the approximation where PBH instantaneously evaporate, $\Tev$ corresponds to the SM temperature just before their sudden evaporation.
The plasma temperature after evaporation could be find by taking into account the PBH entropy injection, and could be computed by the use of Eq.~\eqref{eq:dilution}.

In principle, the $X$ particles produced from the PBH evaporation could decay and produce the observed baryon asymmetry, depending on the Universe conditions when they were emitted. 
Let us assess the maximum amount of asymmetry that could be produced solely from the evaporation, neglecting the effects of the thermal plasma on such generation.
The total number of $X$ particles of mass $M_X$ emitted over the lifetime of a PBH is given by
\begin{equation}
N_{X} = \int_{0}^{\tau}dt\int_{M_{X}}^{\infty}dE\, \frac{d^{2}N_{X}}{dt\, dE} = \int_{M_{\rm BH0}}^{0}d\Mbh\, \frac{30720\pi\, \Mbh^{2}}{\gs\, M_P^{4}} \int_{M_{X}}^{\infty} dE\, \frac{d^{2}N_{X}}{dt\, dE}\, .
\end{equation}
One can solve the above equation in the two limits $M_{X}\ll T_{\rm BH0}$ and $M_{X}\gg T_{\rm BH0}$ where $T_{\rm BH0}\equiv{M_P^{2}}/(8\pi\, M_{{\rm BH0}})$ is the
initial temperature of the PBH. Assuming that $X$ is a fermion and matching the two solutions, we can approximate the total number of $X$ emitted by a single PBH to be
\begin{equation}
 N_{X} \simeq \frac{90\zeta\left(3\right)g_{X}}{\pi^{3}\gs }\times
 \begin{dcases}
 \left(\frac{M_{\rm BH0}}{M_P}\right)^{2} & M_{X}\leq r_{f}\, T_{\rm BH0}\,,\\
 \frac{r_f^2}{64\pi^{2}}\left(\frac{M_P}{M_{X}}\right)^{2} & M_{X}\geq r_{f}\, T_{\rm BH0}\,,
 \end{dcases}\label{eq:N_fermion}
\end{equation}
where $r_{f}\equiv\sqrt{15\zeta\left(5\right)/\zeta\left(3\right)}$. Alternatively, the approximate solution if $X$ is a boson is
\begin{equation}
 N_{X} \simeq \frac{120\zeta\left(3\right)g_{X}}{\pi^{3}\gs } \times
 \begin{dcases}
 \left(\frac{M_{\rm BH0}}{M_P}\right)^{2} & M_{X}\leq r_{b}\, T_{\rm BH0}\,,\\
 \frac{r_b^2}{64\pi^{2}}\left(\frac{M_P}{M_{X}}\right)^{2} & M_{X}\geq r_{b}\, T_{\rm BH0}\,,
 \end{dcases} \label{eq:N_boson}
\end{equation}
where $r_{b}\equiv\sqrt{{12\zeta\left(5\right)/\zeta\left(3\right)}}$.
The total number density of particle
$X$ produced from the evaporation of the PBHs
normalized by cosmic entropy density, $s$, (before taking into account
entropy injection from the evaporation of the PBHs) can be written as
\begin{equation}
 Y_{X}^{0} \equiv \frac{N_{X}\, n\left(\Tev\right)}{s\left(\Tev\right)} = \frac{N_{X}\, n_{0}}{s\left(T_{0}\right)} = \frac{3\beta\, T_{0}\, N_{X}}{4M_{\rm BH0}}\,,\label{eq:Y_i_before}
\end{equation}
where in the final equality, we have used the definition of \equaref{eq:beta} and $\rR\left(T_{0}\right)/s\left(T_{0}\right) = 3\, T_{0}/4$. 

The entropy dilution from the complete evaporation of the PBHs can be estimated using conservation of energy before and after their evaporation\footnote{The additional entropy contribution from the decays of massive particle produced by PBHs with $T_{\rm BH0}\sim m_{i}$ to the radiation will be suppressed by $m_{i}/M_{\rm BH0}$
and hence can be neglected.}
\begin{eqnarray}
 \frac{\pi^{2}}{30}\gs \Tev^{4}+M_{\rm BH0}\frac{n_{0}}{s\left(T_{0}\right)}s\left(\Tev\right) & = & \frac{\pi^{2}}{30}\gs \tilde{T}^{4}\nonumber \\
 \implies \frac{s(\tilde T)}{s(\Tev)} = \left(\frac{\tilde{T}}{\Tev}\right)^{3} & = & \left(1+\frac{\beta T_{0}}{\Tev}\right)^{3/4}\,,
 \label{eq:dilution}
\end{eqnarray}
where $\tilde{T}$ is the temperature of the SM plasma after PBH evaporation occurs.%
\footnote{Alternatively, the entropy dilution can also be computed as in Ref.~\cite{Bernal:2021yyb}.}
If the second factor in the bracket exceeds unity, then the PBHs come to dominate the energy density of the early Universe before they evaporate, resulting in entropy dilution after their complete evaporation.
Here we emphasize that $\tilde T$ is independent of $\beta$, even if PBHs eventually dominate the total energy density of the Universe.
It was shown in e.g. Refs.~\cite{Perez-Gonzalez:2020vnz, Bernal:2020kse, Bernal:2020bjf, Cheek:2021odj, Cheek:2021cfe, Bernal:2021yyb, Bernal:2021bbv} that the entropy injection from PBHs can affect the viable parameter space of baryogenesis and dark matter production, and therefore, such an effect cannot be neglected. Applying the dilution factor of \equaref{eq:dilution} to \equaref{eq:Y_i_before}, we find that
the particle abundance after dilution is
\begin{equation}
 Y_{X} \equiv \frac{3\beta\, T_{0}\, N_{X}}{4M_{\rm BH0}}
\frac{s(\Tev)} {s(\tilde T)}
= \frac{3\beta\, T_{0}\, N_{X}}{4M_{\rm BH0}} \left(1 + \beta\, \frac{T_{0}}{\Tev}\right)^{-3/4}.
\label{eq:Y_i_final}
\end{equation}
Considering a fermionic $X$ with $g_X = 2$, its final abundance from PBH evaporation is
\begin{equation}
 Y_{X} \simeq \frac{135 \zeta(3)}{\pi^{3} \gs}\, \beta \frac{T_{0}}{M_{\rm BH0}}\left(1 + \beta \frac{T_{0}}{\Tev}\right)^{-\frac34}\times
 \begin{dcases}
 \left(\frac{M_{\rm BH0}}{M_P}\right)^{2} & M_{X}\leq r_{f}\, T_{\rm BH0}\,,\\
 \frac{15\zeta\left(5\right)}{64\pi^{2}\zeta\left(3\right)}\left(\frac{M_P}{M_{X}}\right)^{2} & M_{X}\geq r_{f}\, T_{\rm BH0}\,.
 \end{dcases}\label{eq:YN_PBH}
\end{equation}
In general, \equaref{eq:YN_PBH} can be applied since the entropy dilution term is relevant only when the PBHs dominate the cosmic energy density. Note that in the limit of large $\beta\frac{T_0}{T_{\rm ev}} \gg 1$, the $X$ abundance becomes independent of $\beta$. This can be understood as the compensation between the production of $X$ and the entropy dilution from PBH evaporation, which also occurs in the case where $X$ particles are produced from a hotter dark sector~\cite{Bernal:2017zvx}.

\begin{figure}
	\centering
 \includegraphics[scale=0.73]{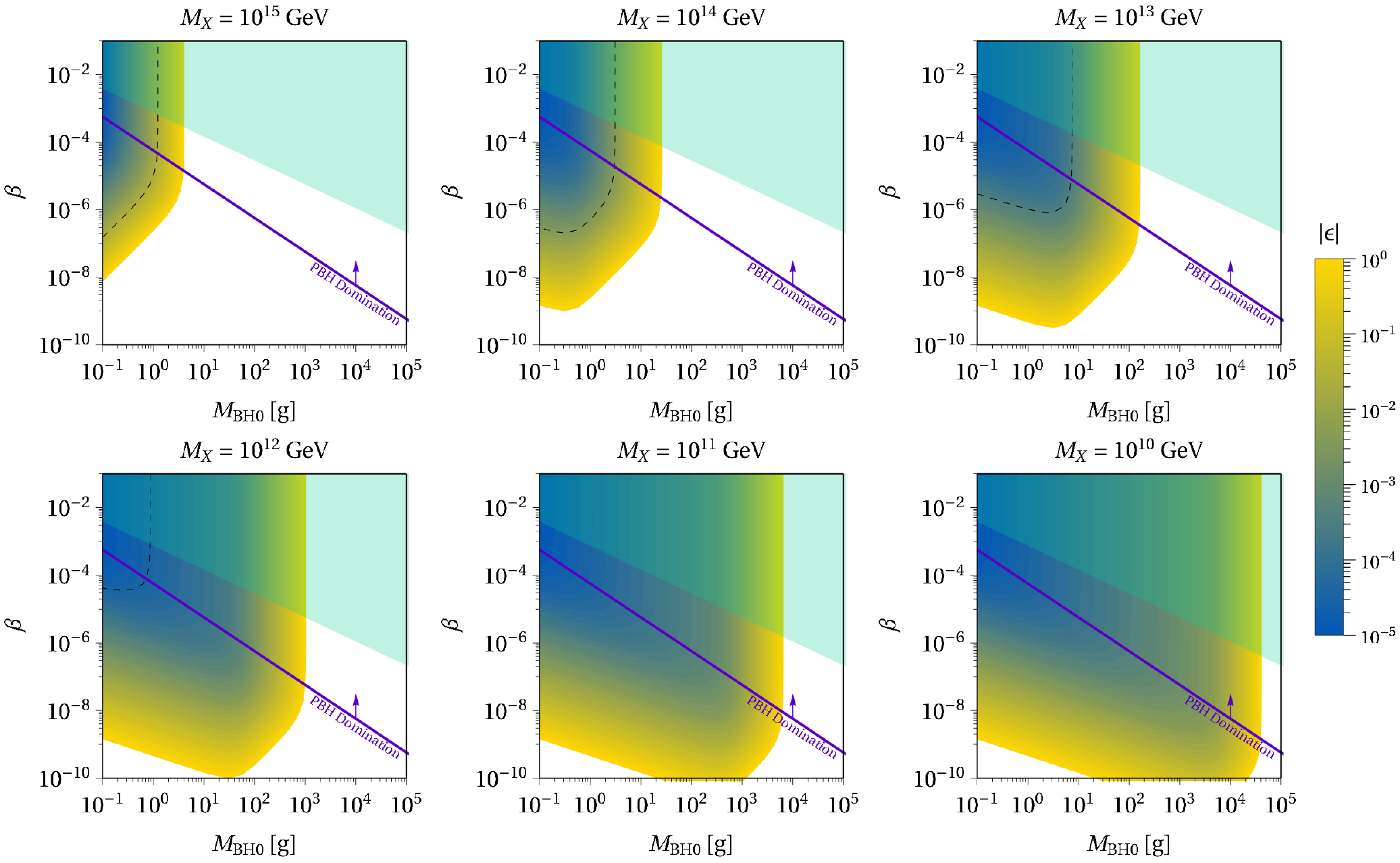}
 \caption{Contours of CP violation parameter $|\epsilon|$ required for $M_X = 10^{15}$, $10^{14}$, $10^{13}$, $10^{12}$, $10^{11}$ and $10^{10}$~GeV (from left to right, top to bottom) taking into account the entropy production from PBH evaporation. The white regions are theoretically excluded while the dashed lines are the Davidson-Ibarra bound for the specific scenario of leptogenesis. For $M_{N_1} \lesssim 10^{12}$~GeV, we require tuning to achieve PBH leptogenesis. The green regions are excluded due GWs.
}
	\label{fig:leptogenesis_PBH}
\end{figure}

Assuming that the decays of $X$ particles violate $B-L$ charge, the maximum baryon number asymmetry normalized by cosmic entropic density that can be obtained when $X$ decays at $T\ll M_{X}$, but before the electroweak sphaleron freezes out, is given by~\cite{Fong:2020fwk}
\begin{equation}
 Y_{B}^{\rm max} = \frac{30}{97}\, Y^{\rm max}_{B-L} =\frac{30}{97}\, \epsilon\, Y_{X}\,,
\label{eq:YBmax}
\end{equation}
where $\epsilon$ is the CP violation parameter for the decay of $X$. 
For the conversion factor 30/97, we have assumed that the electroweak sphaleron freezes out after electroweak symmetry breaking at $T \sim 130$ GeV as indicated by the lattice calculation \cite{DOnofrio:2014rug} and excluded the top quark contribution \cite{Inui:1993wv}.
The scenario above can occur if the $X$ particles are long-lived and/or produced by PBH evaporation at $T \ll M_X$ when all washout processes are entirely suppressed. 
\equaref{eq:YBmax} allows us to determine the minimum required $|\epsilon|$ by imposing such that $Y_B^{\rm max}$ is greater or equal to the observed value $8.7 \times 10^{-11}$~\cite{Planck:2018vyg} and this is plotted in \figref{fig:leptogenesis_PBH} for several values of $M_X$. White regions are excluded as they correspond to $|\epsilon| \geq 1$.
Taking $|\epsilon| = 1$ and requiring successful baryogenesis, we can estimate the upper bound on $M_X$ using the second relation of \equaref{eq:YN_PBH} 
which gives
\begin{equation}
    M_X \lesssim 1.3\times 10^{17}\,{\rm GeV},
    \label{eq:bound_MX}
\end{equation}
where we have chosen $\beta = 10^{-3}$ and $M_{\rm BH0} = 0.1$ g as reference values (in fact, we have saturated the upper bound on $M_X$ for this large value of $\beta$).
The parameter space for viable PBH-assisted baryogenesis increases with decreasing $M_X$ because the production of $X$ from PBH becomes more efficient despite the dilution from entropy production of PBHs. 
Considering the specific case of leptogenesis in the context of type-I seesaw mechanism with a hierarchical mass spectrum of RHNs, the Davidson-Ibarra bound on $|\epsilon|$ for leptogenesis from the lightest RHN, $N_1$, with mass, $M_{N_1}$, is~\cite{Davidson:2002qv}
\begin{equation}
|\epsilon| \leq \frac{3M_{N_1}}{16\pi v^2} \frac{|\Delta m_{\rm atm}^2|}{m_h + m_l}\,,
\label{eq:DI}
\end{equation}
where $|\Delta m_{\rm atm}^2| \equiv m_h^2 - m_l^2$ and $m_h\,(m_l)$ is the heaviest (lightest) light neutrino mass respectively.
This limit is shown in \figref{fig:leptogenesis_PBH} with dashed lines.
Setting $|\Delta m_{\rm atm}^2| = m_h^2 = (0.05~{\rm eV})^2 \gg m_l^2$. For $M_{N_1} \lesssim 10^{12}$~GeV, purely PBH leptogenesis would require tuning to evade the bound of \equaref{eq:DI}, e.g. requiring a quasi-degenerate RHN mass spectrum.
Finally, the green regions are in tension with the GW bound in \equaref{eq:GW}.
\section{Leptogenesis and Primordial Black Holes} \label{sec:lept}

\subsection[High-scale Leptogenesis and $\Delta L=2$ Processes]{\boldmath High-scale Leptogenesis and $\Delta L=2$ Processes} \label{sec:wash}
As alluded to in \secref{sec:intro}, we are interested in high-scale type-I leptogenesis with seesaw scale $M_N \gtrsim 10^{12}$~GeV where there is a sensitivity to the absolute scale of light neutrino mass. At such high scales, due to large neutrino Yukawa coupling, the CP violation required for viable leptogenesis is naturally large without requiring a
quasi-degenerate RHN mass spectrum to enhance the CP violation resonantly. Therefore, we will focus on a mildly hierarchical RHN mass spectrum $M_{N_1} < M_{N_2} < M_{N_3}$, which, together with the strong washout condition, implies that tracking the dynamics of $N_1$ is sufficient as the contributions from $N_2$ and $N_3$ are suppressed. We will also ignore the lepton flavor effects as we consider the high-scale scenario. 
The relevant terms of the type-I seesaw Lagrangian are
\begin{equation}
-{\cal L} \supset \frac{1}{2} M_{N_i} \overline {N_i^c} N_i + \overline{\ell_\alpha} H^* \lambda_{\alpha i} N_i + {\rm H.c.} \,,
\label{eq:lag}
\end{equation}
where we consider three RHNs $N_i$ ($i=1,2,3$), the SM leptonic, $\ell_\alpha$ ($\alpha = e,\mu,\tau$), and Higgs doublet $H$ where the antisymmetric $SU(2)_L$ contractions are implicit.
The CP violation parameter for leptogenesis from $N_1$ decays is defined as:
\begin{equation}
\epsilon \equiv \frac{\Gamma(N_1 \to \ell H) - \Gamma(N_1 \to \bar \ell H^\dagger)}{\Gamma_{N_1}}\,,
\label{eq:cpasym}
\end{equation}
where we have summed over the final lepton flavors and $\Gamma_{N_1}$ is the total $N_1$ decay width given by
\begin{equation}
\Gamma_{N_1} = \frac{(\lambda^\dagger\lambda)_{11}\, M_{N_1}}{8\pi}\,.\label{eq:decaywidth}
\end{equation}
At leading order, the light neutrino mass matrix is given by the seesaw formula
\begin{equation}
m_\nu = -v^2\, \lambda\, M_N^{-1}\, \lambda^T, \label{eq:neutrino_mass}
\end{equation}
where $M_N = {\rm diag}(M_{N_1},M_{N_2},M_{N_3})$. We will denote the light neutrino masses obtained from this matrix to be $m_1$, $m_2$ and $m_3$ where we identify the solar mass splitting as $\Delta m_{\rm sol}^2 = m_2^2 - m_1^2$.

To provide a conservative estimate of the maximum allowed parameter, we must fix certain free parameters that maximize $|\epsilon|$. Since $\Delta m_{\rm sol}^2 \ll |\Delta m_{\rm atm}^2|$, we can approximate $m_1 = m_2$ and using the Casas-Ibarra parametrization~\cite{Casas:2001sr} there is only a single relevant complex angle of the $R$-matrix, $R = R(z_{13})$~\cite{Hambye:2003rt}. Defining $z_{13} \equiv x + i\, y$, where $x$ and $y$ are real, the CP parameter in the hierarchical limit of RHN mass spectrum $M_{N_1} \ll M_{N_2}, M_{N_3}$, is given by
\begin{equation}
|\epsilon| 
= \frac{3M_{N_1}}{16\pi v^2}\, \frac{|\Delta m_{\rm atm}^2|}{m_h + m_l}\,
\frac{\left|\sin(2x)\sinh(2y)\right|}
{\cosh(2y)-f\cos(2x)}\,,
\label{eq:CP1_hie}
\end{equation}
where $f \equiv (m_3 - m_1)/(m_3 + m_1)$. Note that in the denominator, $\cosh(2y) - f \cos(2x) > 0$ given that $\left|f\right| < 1$. In the hierarchical mass limit for light neutrinos, for normal mass ordering $m_3 \gg m_1$ or inverse mass ordering $m_1 \gg m_3$, we have $f \to +1$ or $f \to -1$ respectively, while in the degenerate mass limit $m_3 \sim m_1$, we have $f \ll 1$. 
The maximum $|\epsilon|$ is given by $x = \pm\pi/4$ and $y \to \pm\infty$ which saturates to the Davidson-Ibarra bound of \equaref{eq:DI}. However, as we will see in the following, a large $y$ implies a large washout from inverse decays of $N_1$.  The degree of out-of-equilibrium decay of $N_1$ is quantified by the washout parameter defined as
\begin{equation}
  K \equiv \frac{\Gamma_{N_1}}{H(T = M_{N_1})}
\equiv \frac{\tilde m_1}{m_\star}\,,
\end{equation}
where $H(T=M_{N_1})$ is the Hubble rate in a radiation-dominated Universe evaluated at $T=M_{N_1}$.
In the second definition, $m_\star = \frac{16\pi^2v^2}{3M_P}\sqrt{\frac{g_\star \pi}{5}} \simeq 10^{-3}$~eV while the effective neutrino mass can be expressed as
\begin{equation}
\tilde m_1 
\equiv \frac{(\lambda^\dagger\lambda)_{11}\, v^2}{M_{N_1}}
=\frac{m_h + m_l}{2} \left[ \cosh(2y) - f\cos(2x)\right].
\end{equation}
Since for the heaviest neutrino $m_h > \sqrt{|\Delta m_{\rm atm}^2|} \simeq 0.05$~eV, as long as $|y| > 0.14$, we will always be in the strong washout regime, $K > 1$. 
To determine the upper bound on the neutrino mass scale, $m_h$, the relevant regime is $m_h \gg \sqrt{|\Delta m_{\rm atm}^2|}$ which also implies $m_h + m_l \to 2m_h$ and $f \ll 1$ and therefore
\begin{align}
|\epsilon| &= \frac{3M_{N_1}}{32\pi v^2}
\frac{|\Delta m_{\rm atm}^2\sin(2x)\tanh(2y)|}{m_h}\,, \\
\tilde m_1 &= m_h \cosh(2y)\,.
\end{align}
In this regime, $K > \sqrt{|\Delta m_{\rm atm}^2|}/m_\star \simeq 50$ and hence leptogenesis occurs in the strong washout regime. The CP violation parameter $|\epsilon|$ is maximized for $|x| =\pi/4$ while $y$ should be determined such that $|\epsilon|$ is as large as possible without being overwhelmed by the inverse decay washout controlled $\tilde m_1$. In this regime, an excellent approximation is to maximize $|\epsilon|/\tilde m_1$ which gives $|y| = \frac{1}{2}\log(1+\sqrt{2}) \simeq 0.44$.

In the strong washout regime, the washout processes from inverse decays go out of equilibrium at $T \ll M_{N_1}$ and the asymmetry that survives is produced from decays of $N_1$ below this temperature.
Hence, to determine the upper bound on $m_h$,
we can approximate the total $\Delta L = 2$ washout process with the following squared amplitude valid for $T \ll M_{N_1}$~\cite{Buchmuller:2004nz}
\begin{equation}
 |{\cal A}|^2 = \frac{12\, s}{v^4}\, {\rm Tr}(m_\nu^\dagger m_\nu)\,,
\end{equation}
where $s$ is the center-of-mass energy squared.
The thermally-averaged reaction density for $\Delta L = 2$ processes, assuming Maxwell-Boltzmann statistics, is
\begin{equation}
 \gamma = \frac{T}{64\pi^4}
 \int_0^{\infty} ds\, \sqrt{s}\, {\cal K}_1 \left(\frac{\sqrt{s}}{T}\right) \hat\sigma(s)\,,
\end{equation}
where ${\cal K}_{i}(x)$ is the modified Bessel function of order $i$ and
\begin{equation}
 \hat\sigma(s) = \frac{1}{8\pi s}\int_{-s}^{0} |{\cal A}|^2 dt
 = \frac{3s}{2\pi v^4}{\rm Tr}(m_\nu^\dagger m_\nu)\,.
\end{equation}
After integrating over $s$, we have
\begin{equation}
 \gamma = \frac{3\, T^6}{4\pi^5\, v^4}\,
 {\rm Tr}(m_\nu^\dagger\, m_\nu)\,.
\end{equation}
We note that the interaction above has no Boltzmann suppression even if $T \ll M_{N_1}$ since the $\Delta L = 2$ scatterings ($\bar \ell H^\dagger \leftrightarrow \ell H$ and $\ell \ell \leftrightarrow H^\dagger H^\dagger$) do not involve external $N_1$. Comparing $\gamma/n$ with $n = T^3/\pi^2$ to the Hubble rate, this process will come into thermal equilibrium at a temperature of
\begin{equation}
 T \gtrsim \frac{4\pi^3\, v^4 \times 1.66\sqrt{\gs}}
 {3\, {\rm Tr}(m_\nu^\dagger m_\nu)\, M_P} 
\,,
\end{equation}
assuming a radiation-dominated Universe. It is worth noting that ${\rm Tr}(m_\nu^\dagger m_\nu) = \sum_i m_i^2$ and if this value is $8.5\times 10^{-3}$~eV$^2$ (the current Planck + BOSS bound~\cite{Ivanov:2019hqk}), then the $\Delta L = 2$ scatterings need to be taken into account when $T \gtrsim 6.2\times 10^{12}$~GeV. 
Let us consider leptogenesis which occurs at $T \sim M_{N_1}$ in which $\Delta L = 2$ scatterings are relevant.  
As the $\Delta L = 2$ scatterings are proportional to $m^2_h$, increasing $m_h$ will 
increase the $\Delta L = 2$ scattering rate such that the lepton asymmetry is erased and leptogenesis is no longer viable. We will determine the upper bound on $m_h$ resulting from the washout effect of $\Delta L =2$ scatterings numerically in the next section, with and without assuming the existence of PBHs.

As we are interested in the $m_h$-$M_{N_1}$ parameter space for viable high-scale leptogenesis we can make an estimation of the upper bound on $M_{N_1}$ (without the existence of PBHs): as $M_{N_1}$ is increased, $|\epsilon|$ will also increase. The maximum will be reached before violating perturbativity when $\epsilon\sim \mathcal{O}\left(1\right)$. Considering this upper bound on $\epsilon$ and that the largest $B-L $ asymmetry that can be generated in the strong washout regime is $Y_{B-L} \sim 10^{-4} |\epsilon|$, we can estimate the upper bound on $M_{N_1}$ by restricting the additional exponential washout from $\Delta L = 2$ scatterings to be
\begin{equation}
  \exp\left[-\int_{z_B}^{\infty} dz \frac{\gamma}{z\, H\, n}\right] \lesssim 10^{-6}\,,
\label{eq:upper_bound_MN1}
\end{equation}
where $z \equiv M_{N_1}/T$, and $z_B = M_{N_1}/T_B$ with $T_B$ 
the temperature when the inverse decay rate becomes slower than the Hubble rate. The above restriction is to ensure that after taking into account the additional suppression factor above, we can still generate $Y_{B-L} \gtrsim 10^{-10}$ in accordance with observation~\cite{Planck:2018vyg}.
Considering the least restrictive case $m_h = 0.05$~eV and $z_B \approx 3$~\cite{Buchmuller:2004nz}, we obtain an upper bound $M_{N_1} \lesssim 10^{15}$~GeV.

\subsection{Boltzmann Equations}
In this section, we present the Friedmann and Boltzmann equations that we numerically solve to derive the $m_{h}$ bound including the possible contribution from PBHs. 
The Friedmann equations for the comoving radiation ($\varrho_{\rm R} \equiv a^4 \rR$) and PBHs ($\varrho_{\rm BH} \equiv a^3 \rho_{\rm BH}$) energy densities are
\begin{subequations}\label{eq:UnEv}
\begin{align}
 	aH\frac{d\varrho_{\rm R}}{da} &= -\frac{\varepsilon_{\rm SM}(\Mbh)}{\varepsilon(\Mbh)}\frac{d\ln\Mbh}{dt}a\varrho_{\rm BH}\,,\label{eq:UnEvRad}\\ 
 aH\frac{d\varrho_{\rm BH}}{da} &=\frac{d\ln\Mbh}{dt} \varrho_{\rm BH}\,,\\
 H^2 &=\frac{8\pi}{3\, M_P^2} \left(\varrho_{\rm BH} a^{-3}+\varrho_{\rm R} a^{-4}\right)\,,
\end{align}
\end{subequations}
where $a$ is the scale factor, $H$ the Hubble rate, and $\varepsilon_{\rm SM}(\Mbh)$ contains only the SM contribution to the evaporation. 
Note that we evolve with respect to $a$ instead of $z$, in contrast to standard leptogenesis treatments. 
This is due to the possibly significant entropy dilution present in the PBH scenario.
For convenience, we also track the evolution of the SM thermal plasma temperature, $T$~\cite{Lunardini:2019zob, Bernal:2019lpc, Arias:2019uol}
\begin{equation} \label{eq:TUev}
  aH\frac{dT}{da} = -\frac{T}{\Delta}\left\{H + \frac{\varepsilon_{\rm SM}(\Mbh)}{\varepsilon (\Mbh)}\frac{d\ln\Mbh}{dt} \frac{\gs(T)}{\gss(T)} \frac{a \varrho_{\rm BH}}{4\varrho_{\rm R}}\right\}\,,
\end{equation}
where $\Delta$ takes into account the change on the effective number of entropic degrees of freedom $\gss(T)$ in \equaref{eq:UnEvRad} 
\begin{equation}
\Delta \equiv 1 + \frac{T}{3 \gss(T)}\frac{d\gss(T)}{dT}\,.
\end{equation}
Together with this set of equations, we solve the following momentum-integrated Boltzmann equations for the comoving thermal (${\cal N}_{N_1}^{\rm TH}$) and non-thermal (${\cal N}_{N_1}^{\rm BH}$) RHN densities~\cite{Perez-Gonzalez:2020vnz}
\begin{subequations}\label{eq:BEright-handedn}
\begin{align}
	a H \frac{d{\cal N}_{N_1}^{\rm TH}}{da} &= -({\cal N}_{N_1}^{\rm TH}-{\cal N}_{N_1}^{\rm eq})\Gamma_{N_1}^T\,,\label{eq:BERH-TH}\\
	a H \frac{d{\cal N}_{N_1}^{\rm BH}}{da} &= -{\cal N}_{N_1}^{\rm BH}\Gamma_{N_1}^{\rm BH}+ {\cal N}_{\rm BH} \Gamma_{{\rm BH}\to N_1}\,,\label{eq:BERH-BH}
\end{align}
\end{subequations}
where $\Gamma_{N_1}^T$ and ${\cal N}_{N_1}^{\rm eq}$ are the thermally averaged decay rate and the equilibrium comoving abundance of the RHNs, respectively. 
$\Gamma_{N_1}^{\rm BH}$ in \equaref{eq:BERH-BH} is the decay width corrected by an average inverse time dilatation factor
\begin{equation}\label{eq:GBH}
	\Gamma_{N_1}^{\rm BH} \equiv \left\langle\frac{M_{N_1}}{E_{N_1}}\right\rangle_{\rm BH} \Gamma_{N_1}\,,
\end{equation}
where $\Gamma_{N_1}$ is the RHN decay width defined in \equaref{eq:decaywidth}. 
It is worth emphasizing that the thermal average is taken with respect to the PBH instantaneous spectrum since the RHN energies are distributed according to the Hawking rate, which resembles a thermal distribution. 
To address the generation of RHNs from the PBH density, we have included a source term in
\equaref{eq:BERH-BH} equal to the comoving PBH number density, ${\cal N}_{\rm BH}\equiv \varrho_{\rm BH}/\Mbh$, times $\Gamma_{{\rm BH}\to N_1}$, the total RHN emission rate per BH 
\begin{equation}
  \Gamma_{{\rm BH}\to N_1}=\int_{M_{N_1}}^\infty dE\, \frac{d^{2}N_{N_1}}{dt\, dE}\,.
\end{equation}
The equation for the $B-L$ asymmetry, ${\cal N}_{B-L}$, is given by
\begin{equation}\label{eq:BElep}
	a H \frac{d{\cal N}_{B-L}}{da}= \epsilon\left[({\cal N}_{N_1}^{\rm TH}-{\cal N}_{N_1}^{\rm eq})\Gamma_{N_1}^T + {\cal N}_{N_1}^{\rm BH} \Gamma_{N_1}^{\rm BH}\right] + \left(\frac{1}{2}\Gamma_{N_1}^T {\cal N}_{N_1}^{\rm eq}+\gamma\right)\frac{{\cal N}_{B-L}}{{\cal N}_{\ell}^{\rm eq}}\,,
\end{equation}
where $\epsilon$ is the CP parameter of \equaref{eq:cpasym} describing the decay asymmetry generated by $N_{1}$, and ${\cal N}_{\ell}^{\rm eq}$ is the lepton equilibrium abundance. The term proportional to ${\cal N}_{B-L}$ corresponds to the washout processes including the $\Delta L=2$ interactions discussed in \secref{sec:wash}. 
We compute the baryonic yield from the solution of the Friedmann-Boltzmann equations using
\begin{align}
    Y_{B}=\frac{30}{97}\frac{{\cal N}_{B-L}}{a_{\rm fn}^3s(T_{\rm fn})}\, ,
\end{align}
with $a_{\rm fn}$ and $s(T_{\rm fn})$ are the scale factor and entropy density where we stop the evolution of the Boltzmann and Friedmann equations.

We solve the system of equations \equassref{eq:UnEv}{eq:BEright-handedn}{eq:BElep}, together with the PBH mass rate, \equaref{eq:MEq}, to obtain the final baryon asymmetry.
We make use of {\tt ULYSSES}~\cite{Granelli:2020pim} which not only contains the infrastructure to solve equations for leptogenesis, but also has a library with the BH characteristics for the Schwarzschild and Kerr cases, and fitted forms of the total Hawking emission rate $\Gamma_{{\rm BH}\to N_1}$.
The code used in this work has been made publicly available within {\tt ULYSSES}.\footnote{\url{https://github.com/earlyuniverse/ulysses} \href{https://github.com/earlyuniverse/ulysses}{\faGithub}}

\section{Results and Discussion}\label{sec:results}

As we are interested in obtaining the upper bound on the heaviest active neutrino mass $m_h$, we restrict ourselves to the strong washout regime. In this regime, we fix $|x| = \pi/4$ and $|y| = 0.44$, as detailed in \secref{sec:wash}, and consider a mild hierarchy of the RHN mass spectrum with $M_{N_3} > M_{N_2} > M_{N_1}$. We have verified that as long as the RHN mass spectrum is not quasi-degenerate, the upper bound on $m_h$ is not sensitive to the specific hierarchy of $M_{N_i}$ chosen. The main reason is that the upper bound is determined by $\Delta L = 2$ scatterings, which depend only on $m_h$ and the temperature at which leptogenesis occurs. Since the upper bound on $m_h$ is not very sensitive to the mass hierarchy of $N$'s, we will fix $M_{N_3} = 2 M_{N_2} = 6 M_{N_1}$ where it is a sufficiently good approximation to consider only $N_1$-leptogenesis.

\begin{figure}[t]
\centering
 \includegraphics[width=\textwidth]{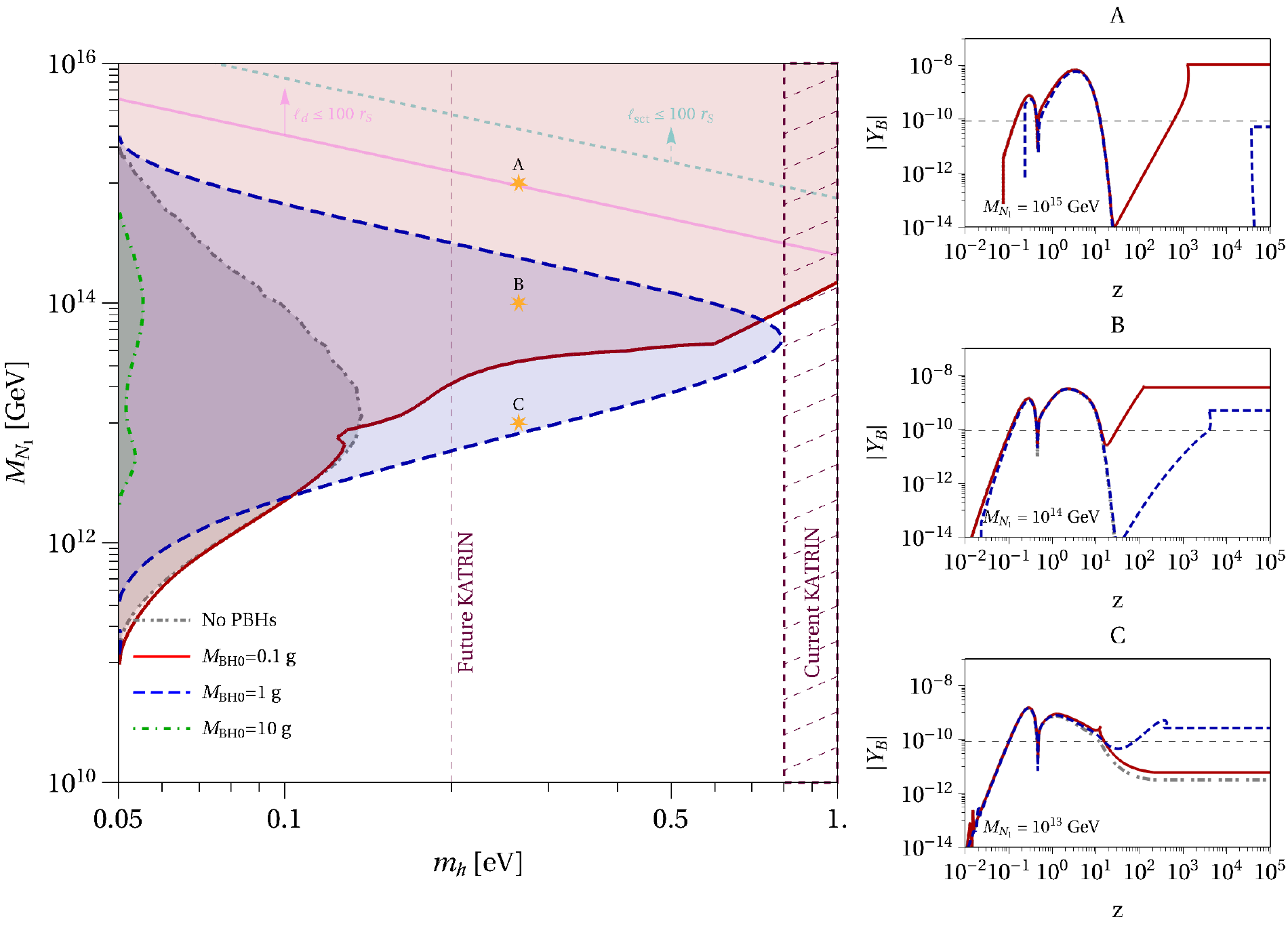}
 \caption{\emph{Left:} Allowed parameter space for leptogenesis in the RH neutrino mass $M_{N_1}$ vs. heaviest active neutrino $m_h$ plane for the standard scenario, i.e. no PBHs (gray dotted region) and including the contribution from evaporating PBHs having initial masses of $M_{\rm BH0}=0.1$~g (red region), $M_{\rm BH0}=1$~g (blue dashed region), $M_{\rm BH0}=10$~g (green dot-dashed region), while fixing $\beta = 10^{-4}$. The dashed region indicates the current KATRIN bound on $m_h$, while the dashed vertical line corresponds to their projected future sensitivity~\cite{KATRIN:2021uub}.  The magenta and cyan lines indicate the region where the $N_1$ decay length $\ell_d$ and mean free path $\ell_s$ are smaller than 100 times the Schwarzschild radius $r_S$, respectively. For such values, we could expect additional washout due to $\Delta L =2$ processes that might be active around the PBH, see text.
 \emph{Right:} Evolution of the baryon yield $|Y_B|$ as function of $z = M_{N_1}/T$ for three different values of the RHN masses ---indicated in the left panel by the stars--- $M_{N_1}=10^{15}$~GeV (top), $M_{N_1}=10^{14}$~GeV (middle), $M_{N_1}=10^{13}$~GeV (bottom), and $m_h=0.27$~eV. The color indicates the different evolution in the presence of a PBH density with masses $M_{\rm BH0}=0.1$~g (red), $M_{\rm BH0}=1$~g (blue dashed), and in the standard case without PBHs (gray dotted).
 \label{fig:res1}}
\end{figure}

In the left panel of \figref{fig:res1}, the colored area on the left of the gray dotted line shows the successful parameter space i.e. $|Y_B| \geq Y_B^{\rm obs}$ in the plane of $m_h$ versus $M_{N_1}$ for this standard scenario regime (i.e., no PBHs).
We observe that the upper bounds of $M_{N_{1}}\sim 10^{15}\,\text{GeV}$ and $m_h \sim 0.1$ eV are due to the sizable $\Delta L=2$ washout as discussed in \secref{sec:wash} which are in agreement with Ref.~\cite{Giudice:2003jh}. The lower bound on  $M_{N_{1}}$ becomes more stringent with increasing $m_h$ due to the Davidson-Ibarra bound, cf. \equaref{eq:DI}.
The purple hatched region corresponds to the current KATRIN bound, while the vertical dashed line is their projected future sensitivity~\cite{KATRIN:2021uub} to the heaviest active neutrino mass, $m_h$.
Next, we consider the additional effect of light PBHs on the $m_h$ and seesaw scale parameter space. In the left panel of \figref{fig:res1}, we also overlay the successful parameter space for PBH-assisted scenario with $\beta = 10^{-4}$, and different PBHs masses: $M_\text{BH0} = 0.1$~g (solid red), 1~g (dashed blue) and 10~g (dash-dotted green).  For $M_\text{BH0} = 0.1$~g, we observe that the viable parameter space is significantly enlarged such that even $m_h\sim 0.5$~eV and $M_{N_{1}}\sim 10^{16}$~GeV provides successful leptogenesis. 

In fact, the upper bound on $M_{N_1}$ can go beyond GUT scale up to \equaref{eq:bound_MX} where $N_1$ particles are too heavy to be efficiently produced by PBH evaporation.
For $M_\text{BH0} = 1$~g, the viable parameter space is smaller than for $M_\text{BH0} = 0.1$~g because the heavier the PBHs, the lower their initial temperature and therefore the less efficient they are at producing heavy RHNs via Hawking evaporation. Nonetheless, even PBHs with a gram-scale mass can significantly enlarge the viable parameter space.
Finally, we observe that for $M_\text{BH0} = 10$~g, shown in green around $M_{N_1} \sim 10^{12}$~GeV, the viable parameter space shrinks compared to the PBH-less leptogenesis scenario. 
This occurs because the heavier PBHs are less efficient at producing such massive RHNs but they still provide sizable entropy injections into the early Universe plasma, diluting the baryon asymmetry produced from thermal leptogenesis. 
This tension between $\gtrsim \mathcal{O}(10)$~g PBHs and thermal leptogenesis has been discussed in detail in Ref.~\cite{Perez-Gonzalez:2020vnz}. 
Interestingly, the successful region in this case presents a reduction around $M_{N_1}\sim 1.5\times 10^{13}$~GeV and $m_h\sim 0.55$~eV. Such reduction appears because the entropy injection depletes the asymmetry produced thermally, while the RHNs produced from the evaporation are not sufficient to generate the observed baryon yield. On the other hand, for $M_{N_1}\lesssim 1.5\times 10^{13}$~GeV, the PBHs emit efficiently RHNs which mitigate the dilution. For $M_{N_1}\gtrsim 1.5\times 10^{13}$~GeV, the reduction due to the entropy from the evaporation is not strong enough to diminish the baryon asymmetry below the observed value.

Even though we expect that washout process to be switched off in the thermal SM bath, the particles produced during the evaporation should, in principle, heat up the plasma around the PBHs~\cite{Das:2021wei}. 
Thus, $\Delta L = 2$ processes might be active in the PBH vicinity, generating a washout of the final asymmetry.
Although a complete determination of the washout in the PBH proximity lies beyond the scope of this paper, we can estimate whether the RH neutrino is able to escape the PBH before decaying. 
Considering all $2\to 2$ processes for $N_1$ scattering~\cite{Hernandez:2016kel}, and taking the plasma temperature to be the Hawking temperature, we can estimate the $N_1$ mean free path as
\begin{align}
 \ell_{s} = \frac{4\pi}{\langle\widetilde{\gamma}_N^{(0)}\rangle}\frac{v^2}{\tilde{m}_1M_{N_1}}r_S \simeq 1.5\times 10^3\, r_S\,  \left(\frac{\rm 0.5~eV}{m_h}\right)\left(\frac{\rm 10^{14}~GeV}{M_{N_1}}\right),
\end{align}
where $\langle\widetilde{\gamma}_N^{(0)}\rangle \equiv \langle\gamma_N^{(0)}\rangle/T$, $\langle\gamma_N^{(0)}\rangle$ being the momentum-averaged $N_1$ scattering rate for vanishing leptonic chemical potential, taken from Ref.~\cite{Hernandez:2016kel}.
Similarly, the decay length is
\begin{align}
    \ell_d &= 8\pi \langle c \rangle \frac{v^2}{\tilde{m}_1M_{N_1}^3} T = 
    \begin{dcases}
        \frac{\langle c \rangle}{2} \frac{v^2}{\tilde{m}_1M_{N_1}^3}\frac{M_P^4}{M_{\rm BH0}^2} r_S &\text{ for } T_{\rm BH0} \geq M_{N_1} \\
        16\pi^2 \langle c \rangle\frac{v^2}{\tilde{m}_1M_{N_1}} r_S &\text{ for } T_{\rm BH0} <  M_{N_1} 
    \end{dcases},
    \notag\\
    &\simeq  
    \begin{dcases}
        10\, r_S \, \left(\frac{\rm 0.5~eV}{m_h}\right) \left(\frac{\rm 10^{14}~GeV}{M_{N_1}}\right)^3 \left(\frac{\rm 1~g}{M_{\rm BH0}}\right)^2  &\text{ for }  T_{\rm BH0} \geq M_{N_1} \\
        500\, r_S\, \left(\frac{\rm 0.5~eV}{m_h}\right) \left(\frac{\rm 10^{14}~GeV}{M_{N_1}}\right) &\text{ for } T_{\rm BH0} <  M_{N_1}
    \end{dcases},
\end{align}
where 
\begin{align}
    \langle c \rangle = \left\langle \frac{M_{N_1}}{E} \right\rangle^{-1}_{\rm BH} \frac{T_{\rm BH}}{M_{N_1}}\simeq 3.72
\end{align}
and $\langle M_{N_1}/E \rangle_{\rm BH}$ is the thermally averaged time dilation factor for the $N_1$ decay, whose value has been determined numerically considering the full Hawking emission spectrum~\cite{Perez-Gonzalez:2020vnz}.
The previous estimations assume that the $N_1$ are produced at the horizon, even though the localization of the emitted particle at emission is meaningless since its Compton wavelength is of the same order as the Schwarzschild horizon.
Thus, in a sense, our estimates are conservative. 
We present in Fig.~\ref{fig:res1} the region of the parameter space where $\ell_s\leq 100\, r_S$ and $\ell_d\leq 100\, r_S$, bounded by the cyan and magenta lines, respectively.
For the decay length, we have that the relevant solution corresponds to the case where $T_{\rm BH0} < M_{N_1}$.
From this, we observe that the washout would affect mainly the asymmetry produced for $M_{\rm BH0} = 0.1$ g, while we do not expect a significant modification for larger initial PBH masses.
We leave a detailed discussion of the washout around a PBH for future work.

In the right panels of \figref{fig:res1}, we show the solutions to the Boltzmann and Friedmann equations for three points in the parameter space for fixed heaviest neutrino mass $m_h = 0.27$~eV and varying RHN masses: $M_{N_{1}} = 10^{15}$~GeV (point $A$), $10^{14}$~GeV ($B$) and $10^{13}$~GeV ($C$). For point $A$,  we observe that $Y_B$ follows the thermal (gray dashed) solution until $z\sim 10$. At this point, light PBH evaporation starts to contribute effectively and produces RHNs which decay and generate a baryon asymmetry. The rapid change in gradient at $z \sim 10^3$ corresponds to the explosive evaporation that occurs at the end of the $0.1$~g PBHs' lifetime. The same pattern is observed for PBHs of mass $1$~g (dashed blue) but the evaporation is later, $z\sim 5\times 10^4$, as the PBHs are heavier. Moreover, the final baryonic yield is lower since these heavier PBHs produce less RHNs. Finally, the purely thermal case (dashed gray) exhibits an exponential decrease in the baryonic yield at $z\sim 10$ due to the $\Delta L =2$ washout being in thermal equilibrium. The behavior for point $B$ is much the same as point $A$. For point $C$ we observe that the evolution of the baryonic yield for the $0.1$~g mass follows that of the thermal case, and it is the PBHs of mass $1$~g that yield the larger baryon asymmetry. This occurs because the lighter PBHs, with $0.1$~g mass, produce the RHNs with this lower mass scale earlier in their evolution when the $\Delta L=2$ washout processes are still active. For $M_\text{BH0} = 10$~g the baryonic yield is so small that it is not shown on these plots.
\begin{figure}
\centering
\includegraphics[width=\textwidth]{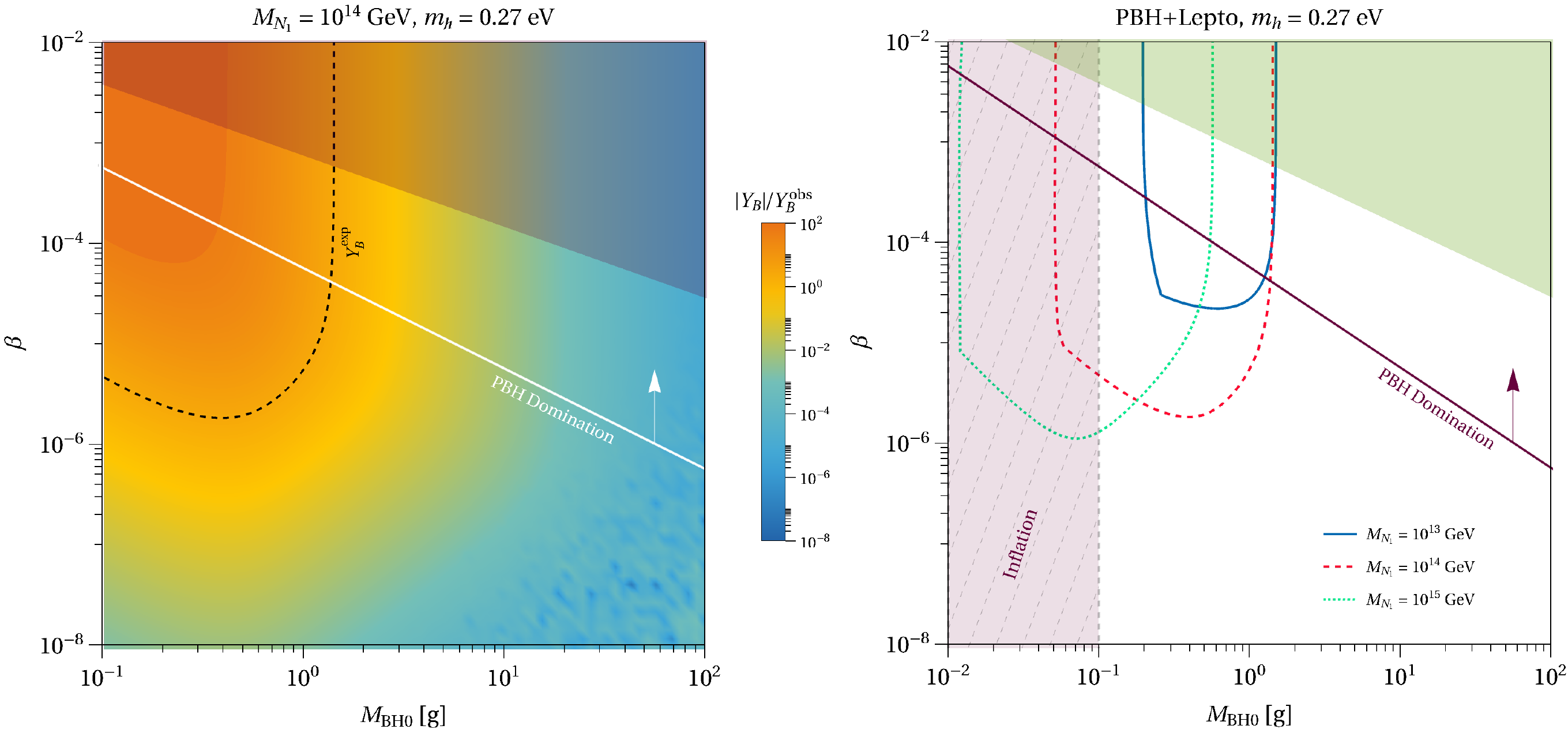}
\caption{{\it Left:} $Y_{B}$ normalized to the observed value $Y_B^{\rm obs}$ for different values of the initial PBH density fraction $\beta$ and mass $M_{\rm BH0}$, taking $M_{N_1}=10^{14}$~GeV and heaviest neutrino mass $m_{h}=0.27$~eV. The dashed line corresponds to the region where we reproduce the measured baryon asymmetry. The white line indicates the PBH parameters which lead to a PBH-dominated era, and the darker region is in tension with GW observations.
{\it Right:} Initial PBH fraction $\beta$ and mass $M_{\rm BH0}$ parameters that generate the measured baryon yield for different values of the RHN masses, $M_{N_1}=10^{13}$~GeV (blue), $M_{N_1}=10^{14}$~GeV (red dashed), and $M_{N_1}=10^{15}$~GeV (emerald dotted). The purple shaded region is excluded from inflation while the green region corresponds to the limit from GWs. The diagonal  purple line indicates the parameters that lead to an early-PBH domination.
\label{fig:res2}}
\end{figure} 
 
In the left panel of \figref{fig:res2}, we plot $Y_{B}$ with PBH-assisted leptogenesis normalized to the observed value $Y_B^{\rm obs}$ for a fixed $M_{N_{1}}=10^{14}$~GeV. The more orange/yellow (green/blue) regions show an enhancement (depletion) of the baryon asymmetry from PBHs. The area to the left of the dashed black line shows the region where the baryonic yield is equal to greater than the measured value.
We observe that for this heavy RHN mass scale, gram or sub-gram scale PBHs ($\beta \gtrsim 10^{-6}$) are required for viable leptogenesis. 
Additionally, in the right panel of \figref{fig:res2} we show the same range in $\beta$ and $M_{\rm BH0}$ for several masses of RHNs: $M_{N_{1}} = 10^{13}$~GeV (blue), $10^{14}$~GeV (red dashed) and $10^{15}$~GeV (emerald dotted), where the regions inside the colored lines are compatible with the observed baryonic yield. The purple (green) colored region is excluded by inflation (GWs). 
We note that the viable parameter space for larger $N_1$ masses shifts towards lighter PBH masses since the lighter PBHs are hotter initially and more efficient in producing heavier $N_1$. The lower bound on the PBH mass is due to too early production of $N_1$ at the temperature when the $\Delta L = 2$ washout is still efficient while the upper bound on the PBH mass is due to the suppression of the production of $N_1$ because heavier PBHs are too cold to produce RHNs but still provide significant entropy production through their evaporation in the PBH domination regime. The viable parameter space shrinks for lower $N_1$ masses due to the Davidson-Ibarra bound in \equaref{eq:DI}.

\section{Conclusions} \label{sec:conclusions}
This work reevaluates viable high-scale thermal leptogenesis parameter space (in terms of the lightest RHN mass $M_1$, and the heaviest active neutrino mass, $m_h$) in the presence of light PBHs. 
In the standard radiation dominated early Universe, if the lightest RHN mass exceeds $\sim 10^{15}\,{\rm GeV}$ and heaviest active neutrino mass is greater than $\sim 0.1$~eV, $\Delta L = 2$ washout processes erase the lepton asymmetry. 
We have demonstrated that the presence of $\lesssim \mathcal{O}(1)\,{\rm g}$ PBHs can provide a non-thermal source of RHNs produced via Hawking radiation when the SM plasma temperature is significantly lower than the lightest RHN mass. 
As such, the washout processes are ineffective, and a lepton asymmetry can be produced through the decays of these PBH sourced RHNs. 
While the leptogenesis parameter space is large, with 18 real parameters needed to determine the Yukawa matrix, we apply the Casas-Ibarra parametrization, assuming a mildly hierarchical RHN mass spectrum, fixing the leptonic mixing angles, mass squared splitting, and CP-violating phase at their best-fit values from global fit data~\cite{Esteban:2020cvm} and fix the Majorana phases to be CP-conserving. 
The remaining parameters of the Yukawa matrix are fixed to maximize the lepton asymmetry without fine-tuning. 

Applying these conservative assumptions, we numerically solve the relevant Friedmann and Boltzmann equations for PBH-assisted leptogenesis, and our main results can be found in \figref{fig:res1}, where the blue and red lines show the enlarged parameter space due to the light PBHs. 
Due to the presence of $\lesssim \mathcal{O}(1)$~g PBHs, we find that the upper bound on the lightest right-handed neutrino mass is generically given by 
$M_{N_{1}}\lesssim 10^{17}$~GeV up to consideration of the perturbativity of the Yukawa matrix 
or by $M_{N_1}\lesssim \textrm{few}\times 10^{15}$~GeV
considering the possibility of heating of thermal bath in the vicinity of the PBHs.
Furthermore, in this PBH-assisted leptogenesis scenario, the heaviest active neutrino mass, $m_{h}$, can be much larger $0.1$~eV. 
Although this is in tension with the current cosmological bound $\sum_i m_i < 0.16$~eV~\cite{Ivanov:2019hqk}, nonstandard cosmology could allow $m_h \sim 1$~eV~\cite{Alvey:2021xmq}. 
The direct neutrino mass measurement from KATRIN gives $m_h < 0.8$~eV while the final sensitivity could go down to 0.2 eV~\cite{KATRIN:2021uub}, essentially probing all the new parameters allowed by PBH-assisted leptogenesis. 
To summarize, the expansion of the viable parameter space (in the $M_{N_{1}}$-$m_h$ plane) implies that high-scale  leptogenesis, excluded in a standard cosmology if neutrino masses are measured to be large ($\gtrsim 0.1$~eV), could be rescued if  gram-scale PBHs once constituted a sizable fraction of the energy density of the Universe.

\section*{Acknowledgments}
NB received funding from the Spanish FEDER/MCIU-AEI under grant FPA2017-84543-P, and the Patrimonio Autónomo - Fondo Nacional de Financiamiento para la Ciencia, la Tecnología y la Innovación Francisco José de Caldas (MinCiencias - Colombia) grants 80740-465-2020 and 80740-492-2021. CSF acknowledges the support by FAPESP Grant No. 2019/11197-6 for the project ``Precision baryogenesis'' and CNPq Grant No. 301271/2019-4.
This work used the DiRAC@Durham facility managed by the Institute for Computational Cosmology on behalf of the STFC DiRAC HPC Facility (\href{www.dirac.ac.uk}{www.dirac.ac.uk}). The equipment was funded by BEIS capital funding via STFC capital grants ST/P002293/1, ST/R002371/1 and ST/S002502/1, Durham University and STFC operations grant ST/R000832/1. DiRAC is part of the National e-Infrastructure. This work has also used the Hamilton HPC Service of Durham University.
This project has received funding/support from the European Union's Horizon 2020 research and innovation programme under the Marie Skłodowska-Curie grant agreement No 860881-HIDDeN.

\bibliographystyle{JHEP}
\bibliography{biblio}

\providecommand{\href}[2]{#2}\begingroup\raggedright\begin{thebibliography}{10}

\bibitem{DiBari:2008mp}
P.~Di~Bari and A.~Riotto, \emph{{Successful type I Leptogenesis with
  SO(10)-inspired mass relations}},
  \href{https://doi.org/10.1016/j.physletb.2008.12.054}{\emph{Phys. Lett. B}
  {\bfseries 671} (2009) 462}
  [\href{https://arxiv.org/abs/0809.2285}{{\ttfamily 0809.2285}}].

\bibitem{Fong:2014gea}
C.S.~Fong, D.~Meloni, A.~Meroni and E.~Nardi, \emph{{Leptogenesis in SO(10)}},
  \href{https://doi.org/10.1007/JHEP01(2015)111}{\emph{JHEP} {\bfseries 01}
  (2015) 111} [\href{https://arxiv.org/abs/1412.4776}{{\ttfamily 1412.4776}}].

\bibitem{King:2021gmj}
S.F.~King, S.~Pascoli, J.~Turner and Y.-L.~Zhou, \emph{{Confronting SO(10) GUTs
  with proton decay and gravitational waves}},
  \href{https://doi.org/10.1007/JHEP10(2021)225}{\emph{JHEP} {\bfseries 10}
  (2021) 225} [\href{https://arxiv.org/abs/2106.15634}{{\ttfamily
  2106.15634}}].

\bibitem{Davidson:2002qv}
S.~Davidson and A.~Ibarra, \emph{{A Lower bound on the right-handed neutrino
  mass from leptogenesis}},
  \href{https://doi.org/10.1016/S0370-2693(02)01735-5}{\emph{Phys. Lett. B}
  {\bfseries 535} (2002) 25}
  [\href{https://arxiv.org/abs/hep-ph/0202239}{{\ttfamily hep-ph/0202239}}].

\bibitem{Buchmuller:2004nz}
W.~Buchmuller, P.~Di~Bari and M.~Plumacher, \emph{{Leptogenesis for
  pedestrians}}, \href{https://doi.org/10.1016/j.aop.2004.02.003}{\emph{Annals
  Phys.} {\bfseries 315} (2005) 305}
  [\href{https://arxiv.org/abs/hep-ph/0401240}{{\ttfamily hep-ph/0401240}}].

\bibitem{Giudice:2003jh}
G.F.~Giudice, A.~Notari, M.~Raidal, A.~Riotto and A.~Strumia, \emph{{Towards a
  complete theory of thermal leptogenesis in the SM and MSSM}},
  \href{https://doi.org/10.1016/j.nuclphysb.2004.02.019}{\emph{Nucl. Phys. B}
  {\bfseries 685} (2004) 89}
  [\href{https://arxiv.org/abs/hep-ph/0310123}{{\ttfamily hep-ph/0310123}}].

\bibitem{Ivanov:2019hqk}
M.M.~Ivanov, M.~Simonovi\'c and M.~Zaldarriaga, \emph{{Cosmological Parameters
  and Neutrino Masses from the Final Planck and Full-Shape BOSS Data}},
  \href{https://doi.org/10.1103/PhysRevD.101.083504}{\emph{Phys. Rev. D}
  {\bfseries 101} (2020) 083504}
  [\href{https://arxiv.org/abs/1912.08208}{{\ttfamily 1912.08208}}].

\bibitem{Allahverdi:2020bys}
R.~Allahverdi et~al., \emph{{The First Three Seconds: a Review of Possible
  Expansion Histories of the Early Universe}},
  \href{https://doi.org/10.21105/astro.2006.16182}{\emph{Open J. Astrophys.}
  {\bfseries 4} (2020) } [\href{https://arxiv.org/abs/2006.16182}{{\ttfamily
  2006.16182}}].

\bibitem{Alvey:2021xmq}
J.~Alvey, M.~Escudero, N.~Sabti and T.~Schwetz, \emph{{Cosmic neutrino
  background detection in large-neutrino-mass cosmologies}},
  \href{https://doi.org/10.1103/PhysRevD.105.063501}{\emph{Phys. Rev. D}
  {\bfseries 105} (2022) 063501}
  [\href{https://arxiv.org/abs/2111.14870}{{\ttfamily 2111.14870}}].

\bibitem{Hawking:1975vcx}
S.W.~Hawking, \emph{{Particle Creation by Black Holes}},
  \href{https://doi.org/10.1007/BF02345020}{\emph{Commun. Math. Phys.}
  {\bfseries 43} (1975) 199}.

\bibitem{Toussaint:1978br}
D.~Toussaint, S.B.~Treiman, F.~Wilczek and A.~Zee, \emph{{Matter - Antimatter
  Accounting, Thermodynamics, and Black Hole Radiation}},
  \href{https://doi.org/10.1103/PhysRevD.19.1036}{\emph{Phys. Rev. D}
  {\bfseries 19} (1979) 1036}.

\bibitem{Hawking:1980ng}
S.W.~Hawking, \emph{{Interacting Quantum Fields Around a Black Hole}},
  \href{https://doi.org/10.1007/BF01208279}{\emph{Commun. Math. Phys.}
  {\bfseries 80} (1981) 421}.

\bibitem{Barrow:1990he}
J.D.~Barrow, E.J.~Copeland, E.W.~Kolb and A.R.~Liddle, \emph{{Baryogenesis in
  extended inflation. 2. Baryogenesis via primordial black holes}},
  \href{https://doi.org/10.1103/PhysRevD.43.984}{\emph{Phys. Rev. D} {\bfseries
  43} (1991) 984}.

\bibitem{Majumdar:1995yr}
A.S.~Majumdar, P.~Das~Gupta and R.P.~Saxena, \emph{{Baryogenesis from black
  hole evaporation}},
  \href{https://doi.org/10.1142/S0218271895000363}{\emph{Int. J. Mod. Phys. D}
  {\bfseries 4} (1995) 517}.

\bibitem{Bugaev:2001xr}
E.V.~Bugaev, M.G.~Elbakidze and K.V.~Konishchev, \emph{{Baryon asymmetry of the
  universe from evaporation of primordial black holes}},
  \href{https://doi.org/10.1134/1.1563709}{\emph{Phys. Atom. Nucl.} {\bfseries
  66} (2003) 476} [\href{https://arxiv.org/abs/astro-ph/0110660}{{\ttfamily
  astro-ph/0110660}}].

\bibitem{Baumann:2007yr}
D.~Baumann, P.J.~Steinhardt and N.~Turok, \emph{{Primordial Black Hole
  Baryogenesis}},  \href{https://arxiv.org/abs/hep-th/0703250}{{\ttfamily
  hep-th/0703250}}.

\bibitem{Fujita:2014hha}
T.~Fujita, M.~Kawasaki, K.~Harigaya and R.~Matsuda, \emph{{Baryon asymmetry,
  dark matter, and density perturbation from primordial black holes}},
  \href{https://doi.org/10.1103/PhysRevD.89.103501}{\emph{Phys. Rev. D}
  {\bfseries 89} (2014) 103501}
  [\href{https://arxiv.org/abs/1401.1909}{{\ttfamily 1401.1909}}].

\bibitem{Morrison:2018xla}
L.~Morrison, S.~Profumo and Y.~Yu, \emph{{Melanopogenesis: Dark Matter of
  (almost) any Mass and Baryonic Matter from the Evaporation of Primordial
  Black Holes weighing a Ton (or less)}},
  \href{https://doi.org/10.1088/1475-7516/2019/05/005}{\emph{JCAP} {\bfseries
  05} (2019) 005} [\href{https://arxiv.org/abs/1812.10606}{{\ttfamily
  1812.10606}}].

\bibitem{Ambrosone:2021lsx}
A.~Ambrosone, R.~Calabrese, D.F.G.~Fiorillo, G.~Miele and S.~Morisi,
  \emph{{Towards baryogenesis via absorption from primordial black holes}},
  \href{https://doi.org/10.1103/PhysRevD.105.045001}{\emph{Phys. Rev. D}
  {\bfseries 105} (2022) 045001}
  [\href{https://arxiv.org/abs/2106.11980}{{\ttfamily 2106.11980}}].

\bibitem{Hooper:2020otu}
D.~Hooper and G.~Krnjaic, \emph{{GUT Baryogenesis With Primordial Black
  Holes}}, \href{https://doi.org/10.1103/PhysRevD.103.043504}{\emph{Phys. Rev.
  D} {\bfseries 103} (2021) 043504}
  [\href{https://arxiv.org/abs/2010.01134}{{\ttfamily 2010.01134}}].

\bibitem{Perez-Gonzalez:2020vnz}
Y.F.~Perez-Gonzalez and J.~Turner, \emph{{Assessing the tension between a black
  hole dominated early universe and leptogenesis}},
  \href{https://doi.org/10.1103/PhysRevD.104.103021}{\emph{Phys. Rev. D}
  {\bfseries 104} (2021) 103021}
  [\href{https://arxiv.org/abs/2010.03565}{{\ttfamily 2010.03565}}].

\bibitem{Datta:2020bht}
S.~Datta, A.~Ghosal and R.~Samanta, \emph{{Baryogenesis from ultralight
  primordial black holes and strong gravitational waves from cosmic strings}},
  \href{https://doi.org/10.1088/1475-7516/2021/08/021}{\emph{JCAP} {\bfseries
  08} (2021) 021} [\href{https://arxiv.org/abs/2012.14981}{{\ttfamily
  2012.14981}}].

\bibitem{JyotiDas:2021shi}
S.~Jyoti~Das, D.~Mahanta and D.~Borah, \emph{{Low scale leptogenesis and dark
  matter in the presence of primordial black holes}},
  \href{https://doi.org/10.1088/1475-7516/2021/11/019}{\emph{JCAP} {\bfseries
  11} (2021) 019} [\href{https://arxiv.org/abs/2104.14496}{{\ttfamily
  2104.14496}}].

\bibitem{Barman:2021ost}
B.~Barman, D.~Borah, S.J.~Das and R.~Roshan, \emph{{Non-thermal origin of
  asymmetric dark matter from inflaton and primordial black holes}},
  \href{https://doi.org/10.1088/1475-7516/2022/03/031}{\emph{JCAP} {\bfseries
  03} (2022) 031} [\href{https://arxiv.org/abs/2111.08034}{{\ttfamily
  2111.08034}}].

\bibitem{Giudice:2000ex}
G.F.~Giudice, E.W.~Kolb and A.~Riotto, \emph{{Largest temperature of the
  radiation era and its cosmological implications}},
  \href{https://doi.org/10.1103/PhysRevD.64.023508}{\emph{Phys. Rev. D}
  {\bfseries 64} (2001) 023508}
  [\href{https://arxiv.org/abs/hep-ph/0005123}{{\ttfamily hep-ph/0005123}}].

\bibitem{Bernal:2021kaj}
N.~Bernal and C.S.~Fong, \emph{{Dark matter and leptogenesis from gravitational
  production}},
  \href{https://doi.org/10.1088/1475-7516/2021/06/028}{\emph{JCAP} {\bfseries
  06} (2021) 028} [\href{https://arxiv.org/abs/2103.06896}{{\ttfamily
  2103.06896}}].

\bibitem{KATRIN:2021uub}
{\scshape KATRIN} collaboration, \emph{{Direct neutrino-mass measurement with
  sub-electronvolt sensitivity}},
  \href{https://doi.org/10.1038/s41567-021-01463-1}{\emph{Nature Phys.}
  {\bfseries 18} (2022) 160}
  [\href{https://arxiv.org/abs/2105.08533}{{\ttfamily 2105.08533}}].

\bibitem{Carr:1974nx}
B.J.~Carr and S.W.~Hawking, \emph{{Black holes in the early Universe}},
  {\emph{Mon. Not. Roy. Astron. Soc.} {\bfseries 168} (1974) 399}.

\bibitem{Press:1973iz}
W.H.~Press and P.~Schechter, \emph{{Formation of galaxies and clusters of
  galaxies by selfsimilar gravitational condensation}},
  \href{https://doi.org/10.1086/152650}{\emph{Astrophys. J.} {\bfseries 187}
  (1974) 425}.

\bibitem{Carr:2009jm}
B.J.~Carr, K.~Kohri, Y.~Sendouda and J.~Yokoyama, \emph{{New cosmological
  constraints on primordial black holes}},
  \href{https://doi.org/10.1103/PhysRevD.81.104019}{\emph{Phys. Rev. D}
  {\bfseries 81} (2010) 104019}
  [\href{https://arxiv.org/abs/0912.5297}{{\ttfamily 0912.5297}}].

\bibitem{Carr:2020gox}
B.~Carr, K.~Kohri, Y.~Sendouda and J.~Yokoyama, \emph{{Constraints on
  primordial black holes}},
  \href{https://doi.org/10.1088/1361-6633/ac1e31}{\emph{Rept. Prog. Phys.}
  {\bfseries 84} (2021) 116902}
  [\href{https://arxiv.org/abs/2002.12778}{{\ttfamily 2002.12778}}].

\bibitem{Planck:2018jri}
{\scshape Planck} collaboration, \emph{{Planck 2018 results. X. Constraints on
  inflation}}, \href{https://doi.org/10.1051/0004-6361/201833887}{\emph{Astron.
  Astrophys.} {\bfseries 641} (2020) A10}
  [\href{https://arxiv.org/abs/1807.06211}{{\ttfamily 1807.06211}}].

\bibitem{Rasanen:2018fom}
S.~Rasanen and E.~Tomberg, \emph{{Planck scale black hole dark matter from
  Higgs inflation}},
  \href{https://doi.org/10.1088/1475-7516/2019/01/038}{\emph{JCAP} {\bfseries
  01} (2019) 038} [\href{https://arxiv.org/abs/1810.12608}{{\ttfamily
  1810.12608}}].

\bibitem{Sasaki:2018dmp}
M.~Sasaki, T.~Suyama, T.~Tanaka and S.~Yokoyama, \emph{{Primordial black
  holes\textemdash{}perspectives in gravitational wave astronomy}},
  \href{https://doi.org/10.1088/1361-6382/aaa7b4}{\emph{Class. Quant. Grav.}
  {\bfseries 35} (2018) 063001}
  [\href{https://arxiv.org/abs/1801.05235}{{\ttfamily 1801.05235}}].

\bibitem{Carr:2020xqk}
B.~Carr and F.~Kuhnel, \emph{{Primordial Black Holes as Dark Matter: Recent
  Developments}},
  \href{https://doi.org/10.1146/annurev-nucl-050520-125911}{\emph{Ann. Rev.
  Nucl. Part. Sci.} {\bfseries 70} (2020) 355}
  [\href{https://arxiv.org/abs/2006.02838}{{\ttfamily 2006.02838}}].

\bibitem{MacGibbon:1990zk}
J.H.~MacGibbon and B.R.~Webber, \emph{{Quark and gluon jet emission from
  primordial black holes: The instantaneous spectra}},
  \href{https://doi.org/10.1103/PhysRevD.41.3052}{\emph{Phys. Rev.} {\bfseries
  D41} (1990) 3052}.

\bibitem{MacGibbon:1991tj}
J.H.~MacGibbon, \emph{{Quark and gluon jet emission from primordial black
  holes. 2. The Lifetime emission}},
  \href{https://doi.org/10.1103/PhysRevD.44.376}{\emph{Phys. Rev.} {\bfseries
  D44} (1991) 376}.

\bibitem{Cheek:2021odj}
A.~Cheek, L.~Heurtier, Y.F.~Perez-Gonzalez and J.~Turner, \emph{{Primordial
  black hole evaporation and dark matter production. I. Solely Hawking
  radiation}}, \href{https://doi.org/10.1103/PhysRevD.105.015022}{\emph{Phys.
  Rev. D} {\bfseries 105} (2022) 015022}
  [\href{https://arxiv.org/abs/2107.00013}{{\ttfamily 2107.00013}}].

\bibitem{Lunardini:2019zob}
C.~Lunardini and Y.F.~Perez-Gonzalez, \emph{{Dirac and Majorana neutrino
  signatures of primordial black holes}},
  \href{https://doi.org/10.1088/1475-7516/2020/08/014}{\emph{JCAP} {\bfseries
  08} (2020) 014} [\href{https://arxiv.org/abs/1910.07864}{{\ttfamily
  1910.07864}}].

\bibitem{Papanikolaou:2020qtd}
T.~Papanikolaou, V.~Vennin and D.~Langlois, \emph{{Gravitational waves from a
  universe filled with primordial black holes}},
  \href{https://doi.org/10.1088/1475-7516/2021/03/053}{\emph{JCAP} {\bfseries
  03} (2021) 053} [\href{https://arxiv.org/abs/2010.11573}{{\ttfamily
  2010.11573}}].

\bibitem{Domenech:2020ssp}
G.~Dom\`enech, C.~Lin and M.~Sasaki, \emph{{Gravitational wave constraints on
  the primordial black hole dominated early universe}},
  \href{https://doi.org/10.1088/1475-7516/2021/04/062}{\emph{JCAP} {\bfseries
  04} (2021) 062} [\href{https://arxiv.org/abs/2012.08151}{{\ttfamily
  2012.08151}}].

\bibitem{Bernal:2021yyb}
N.~Bernal, F.~Hajkarim and Y.~Xu, \emph{{Axion Dark Matter in the Time of
  Primordial Black Holes}},
  \href{https://doi.org/10.1103/PhysRevD.104.075007}{\emph{Phys. Rev. D}
  {\bfseries 104} (2021) 075007}
  [\href{https://arxiv.org/abs/2107.13575}{{\ttfamily 2107.13575}}].

\bibitem{Bernal:2020kse}
N.~Bernal and {\'O}.~Zapata, \emph{{Self-interacting Dark Matter from
  Primordial Black Holes}},
  \href{https://doi.org/10.1088/1475-7516/2021/03/007}{\emph{JCAP} {\bfseries
  03} (2021) 007} [\href{https://arxiv.org/abs/2010.09725}{{\ttfamily
  2010.09725}}].

\bibitem{Bernal:2020bjf}
N.~Bernal and {\'O}.~Zapata, \emph{{Dark Matter in the Time of Primordial Black
  Holes}}, \href{https://doi.org/10.1088/1475-7516/2021/03/015}{\emph{JCAP}
  {\bfseries 03} (2021) 015}
  [\href{https://arxiv.org/abs/2011.12306}{{\ttfamily 2011.12306}}].

\bibitem{Cheek:2021cfe}
A.~Cheek, L.~Heurtier, Y.F.~Perez-Gonzalez and J.~Turner, \emph{{Primordial
  black hole evaporation and dark matter production. II. Interplay with the
  freeze-in or freeze-out mechanism}},
  \href{https://doi.org/10.1103/PhysRevD.105.015023}{\emph{Phys. Rev. D}
  {\bfseries 105} (2022) 015023}
  [\href{https://arxiv.org/abs/2107.00016}{{\ttfamily 2107.00016}}].

\bibitem{Bernal:2021bbv}
N.~Bernal, Y.F.~Perez-Gonzalez, Y.~Xu and {\'O}.~Zapata, \emph{{ALP dark matter
  in a primordial black hole dominated universe}},
  \href{https://doi.org/10.1103/PhysRevD.104.123536}{\emph{Phys. Rev. D}
  {\bfseries 104} (2021) 123536}
  [\href{https://arxiv.org/abs/2110.04312}{{\ttfamily 2110.04312}}].

\bibitem{Bernal:2017zvx}
N.~Bernal and C.S.~Fong, \emph{{Hot Leptogenesis from Thermal Dark Matter}},
  \href{https://doi.org/10.1088/1475-7516/2017/10/042}{\emph{JCAP} {\bfseries
  10} (2017) 042} [\href{https://arxiv.org/abs/1707.02988}{{\ttfamily
  1707.02988}}].

\bibitem{Fong:2020fwk}
C.S.~Fong, \emph{{Baryogenesis in the Standard Model and its Supersymmetric
  Extension}}, \href{https://doi.org/10.1103/PhysRevD.103.L051705}{\emph{Phys.
  Rev. D} {\bfseries 103} (2021) L051705}
  [\href{https://arxiv.org/abs/2012.03973}{{\ttfamily 2012.03973}}].

\bibitem{DOnofrio:2014rug}
M.~D'Onofrio, K.~Rummukainen and A.~Tranberg, \emph{{Sphaleron Rate in the
  Minimal Standard Model}},
  \href{https://doi.org/10.1103/PhysRevLett.113.141602}{\emph{Phys. Rev. Lett.}
  {\bfseries 113} (2014) 141602}
  [\href{https://arxiv.org/abs/1404.3565}{{\ttfamily 1404.3565}}].

\bibitem{Inui:1993wv}
T.~Inui, T.~Ichihara, Y.~Mimura and N.~Sakai, \emph{{Cosmological baryon
  asymmetry in supersymmetric Standard Models and heavy particle effects}},
  \href{https://doi.org/10.1016/0370-2693(94)90031-0}{\emph{Phys. Lett. B}
  {\bfseries 325} (1994) 392}
  [\href{https://arxiv.org/abs/hep-ph/9310268}{{\ttfamily hep-ph/9310268}}].

\bibitem{Planck:2018vyg}
{\scshape Planck} collaboration, \emph{{Planck 2018 results. VI. Cosmological
  parameters}},
  \href{https://doi.org/10.1051/0004-6361/201833910}{\emph{Astron. Astrophys.}
  {\bfseries 641} (2020) A6}
  [\href{https://arxiv.org/abs/1807.06209}{{\ttfamily 1807.06209}}].

\bibitem{Casas:2001sr}
J.A.~Casas and A.~Ibarra, \emph{{Oscillating neutrinos and $\mu \to e,
  \gamma$}}, \href{https://doi.org/10.1016/S0550-3213(01)00475-8}{\emph{Nucl.
  Phys. B} {\bfseries 618} (2001) 171}
  [\href{https://arxiv.org/abs/hep-ph/0103065}{{\ttfamily hep-ph/0103065}}].

\bibitem{Hambye:2003rt}
T.~Hambye, Y.~Lin, A.~Notari, M.~Papucci and A.~Strumia, \emph{{Constraints on
  neutrino masses from leptogenesis models}},
  \href{https://doi.org/10.1016/j.nuclphysb.2004.06.027}{\emph{Nucl. Phys. B}
  {\bfseries 695} (2004) 169}
  [\href{https://arxiv.org/abs/hep-ph/0312203}{{\ttfamily hep-ph/0312203}}].

\bibitem{Bernal:2019lpc}
N.~Bernal and F.~Hajkarim, \emph{{Primordial Gravitational Waves in Nonstandard
  Cosmologies}}, \href{https://doi.org/10.1103/PhysRevD.100.063502}{\emph{Phys.
  Rev. D} {\bfseries 100} (2019) 063502}
  [\href{https://arxiv.org/abs/1905.10410}{{\ttfamily 1905.10410}}].

\bibitem{Arias:2019uol}
P.~Arias, N.~Bernal, A.~Herrera and C.~Maldonado, \emph{{Reconstructing
  Non-standard Cosmologies with Dark Matter}},
  \href{https://doi.org/10.1088/1475-7516/2019/10/047}{\emph{JCAP} {\bfseries
  10} (2019) 047} [\href{https://arxiv.org/abs/1906.04183}{{\ttfamily
  1906.04183}}].

\bibitem{Granelli:2020pim}
A.~Granelli, K.~Moffat, Y.F.~Perez-Gonzalez, H.~Schulz and J.~Turner,
  \emph{{ULYSSES: Universal LeptogeneSiS Equation Solver}},
  \href{https://doi.org/10.1016/j.cpc.2020.107813}{\emph{Comput. Phys. Commun.}
  {\bfseries 262} (2021) 107813}
  [\href{https://arxiv.org/abs/2007.09150}{{\ttfamily 2007.09150}}].

\bibitem{Das:2021wei}
S.~Das and A.~Hook, \emph{{Black hole production of monopoles in the early
  universe}}, \href{https://doi.org/10.1007/JHEP12(2021)145}{\emph{JHEP}
  {\bfseries 12} (2021) 145}
  [\href{https://arxiv.org/abs/2109.00039}{{\ttfamily 2109.00039}}].

\bibitem{Hernandez:2016kel}
P.~Hern\'andez, M.~Kekic, J.~L\'opez-Pav\'on, J.~Racker and J.~Salvado,
  \emph{{Testable Baryogenesis in Seesaw Models}},
  \href{https://doi.org/10.1007/JHEP08(2016)157}{\emph{JHEP} {\bfseries 08}
  (2016) 157} [\href{https://arxiv.org/abs/1606.06719}{{\ttfamily
  1606.06719}}].

\bibitem{Esteban:2020cvm}
I.~Esteban, M.C.~Gonzalez-Garcia, M.~Maltoni, T.~Schwetz and A.~Zhou,
  \emph{{The fate of hints: updated global analysis of three-flavor neutrino
  oscillations}}, \href{https://doi.org/10.1007/JHEP09(2020)178}{\emph{JHEP}
  {\bfseries 09} (2020) 178}
  [\href{https://arxiv.org/abs/2007.14792}{{\ttfamily 2007.14792}}].

\end{thebibliography}\endgroup

\end{document}